\documentclass[preprint,preprintnumbers,superscriptaddress,amsmath]{revtex4}
\usepackage[dvips]{graphics}
\usepackage{epsfig}
\usepackage{psfrag}
\usepackage[psamsfonts]{amssymb}
\usepackage{amsmath}
\usepackage{indentfirst}
\usepackage{amssymb}
\usepackage{wrapfig}
\usepackage{url}

\newtheorem{theorem}{Theorem}

\begin{document}

\title{Hierarchy of Temporal Responses \\ of Multivariate Self-Excited Epidemic Processes}
%The Shorter the Memory of the Kernel,\\ the Longer the Memory of the Response Function\\in Long Memory  Processes}
\author{A. Saichev}
\affiliation{Department of Management, Technology and Economics,
ETH Zurich, Kreuzplatz 5, CH-8032 Zurich, Switzerland}
\affiliation{Mathematical Department,
Nizhny Novgorod State University, Gagarin prosp. 23,
Nizhny Novgorod, 603950, Russia}
\email{saichev@hotmail.com,dsornette@ethz.ch}

\author{D. Sornette}
\affiliation{Department of Management, Technology and Economics,
ETH Zurich, Kreuzplatz 5, CH-8032 Zurich, Switzerland}

\begin{abstract}
We present the first exact analysis of some of the temporal properties
of multivariate self-excited Hawkes conditional Poisson processes, which
constitute powerful representations of a large variety of systems
with bursty events, for which past activity
triggers future activity.  The term ``multivariate'' refers to the property that
events come in different types, with possibly different intra- and inter-triggering
abilities. We develop the general
formalism of the multivariate generating moment function for the cumulative number
of first-generation and of all generation events triggered by a given mother event (the ``shock'')
as a function of the current time $t$. This corresponds to studying the response
function of the process. A variety of different systems have been analyzed.
In particular, for systems in which triggering between events
of different types proceeds through a one-dimension directed
or symmetric chain of influence in type space, we report a novel hierarchy of intermediate
asymptotic power law decays $\sim 1/t^{1-(m+1)\theta}$ of the rate of triggered events as a function
of the distance $m$ of the events to the initial shock in the type space, where $0 < \theta <1$
for the relevant long-memory processes characterizing many natural and social systems.
The richness of the generated time dynamics comes from the
cascades of intermediate events of possibly different kinds, unfolding
via a kind of inter-breeding genealogy.
\end{abstract}

\date{\today}

\maketitle

\section{Introduction}

We study a class of point processes that was introduced by Hawkes in 1971
\cite{Hawkes1,Hawkes2,Hawkes3,Hawkes4}. It
is much richer and relevant to most natural
and social systems than standard point processes
\cite{Bremaud1,DVJ2007,shotnoiseintro,Montroll12,Scher,ContTankov},
because it describes ``self-excited'' processes.  This term means that the past events
have the ability to trigger future events, i.e., $\lambda(t | H_t)$ is a function of past events,
being therefore non-markovian.
Many works have been performed to characterize the statistical and dynamical properties
of this class of models, with applications ranging from geophysical \cite{Ogata88,Ogata1,Ogata2,Ogata3,HS02,SaiSor07},
medical \cite{SorOsorio10} to financial systems,
with applications to Value-at-Risk modeling \cite{Chavezetal05}, high-frequency price processes
\cite{BauwensHautsch09}, portfolio credit risks \cite{Eymanetal10}, cascades
of corporate defaults \cite{Azizetal10}, financial contagion \cite{Aitsahaliaetal10},
and yield curve dynamics \cite{SalmonTham08}.

While surprisingly rich and  powerful in explaining empirical observations in a variety of
systems, most previous studies have used mono-variate self-excited point processes,
i.e., they have assumed the existence of only a single type of events, all the events presenting some
ability to trigger events of the same type.
However, in reality, in many systems,
events come in different types with possibly different properties,
while keeping a degree of mutual inter-excitations.
\begin{itemize}
\item Earthquakes are partitioned
in different tectonic regions.
\item Neuronal excitations in the brain
come in different types, such as spikes, bursts and seizures.
In addition, epileptic seizures involve different brain structures at different scales.
\item Financial volatility bursts occur at a given time on some assets and not on others.
\item Defaults on debts may start and develop preferentially on some firms in some industrial sectors.
\item Some countries within a currency block may start to exhibit specific sovereign risks and not other
countries.
\item Only a subset of the population of bloggers or of developers may be active
at any time.
\end{itemize}
Notwithstanding the existence of such different types, categories or locations,
events of one type may trigger both new events of the same type as well as
of different types.  For instance,
\begin{enumerate}
\item  there is now
convincing evidence of earthquake triggering at large distances
across tectonic plate boundaries.
\item Epileptic seizures are nowadays believed to be
able to trigger large scale neuronal excitations over largely different brain structures.
It is possible that even neuronal spikes, bursts or different types
of subclinical epileptic seizures may play a role in
triggering clinical epileptic seizures of different types in distinct cortical structures.
\item The hectic price movement of a given stock may trigger by contagion large volatility bursts
in other stocks or assets, via technical or behavioral transfer mechanisms.
Financial volatility bursts occurring for one asset may trigger
future volatility fluctuations for other assets.
\item The shocks due to defaults
of some firms may jump across industries to encompass different branches
through a chain reaction.
\item Countries of different regions and currencies may become coupled through
the direct and indirect flows of international commerce as well as mutual debt ownerships.
\item The activity of a blogger or a developer may be encouraged by previous actions of
other agents, leading to future commits triggered by previous commits of different people.
\end{enumerate}

These observations suggest that multivariate self-excited point processes, which extend
the class of mono-variate self-excited point processes, provide a very important
class of models to describe the self-excitation (or  intra-triggering) as well as the mutual influences
or triggering between different types of events that occur in many natural and social systems.
Actually, the generalization to multivariate self-excited point processes was
mentioned by Hawkes himself in his first paper \cite{Hawkes1}, but
the full relevance of this class of models has only been recently appreciated
\cite{Zhang-Ma05,Aitsahaliaetal10}.

The organization of the paper proceeds as follows. Section 2 first recalls the
definition of the monovariate Hawkes process and then presents the general
multivariate self-exciting Hawkes processes.   Section 3 develops the
formalism of the multivariate generating moment function for the cumulative number
of first-generation and of all generation events triggered by a given mother event
as a function of the current time $t$. Section 4 provides the general relations
to obtain the mean numbers of events triggered over all generations
by a given event as a function of time. Section 5 analyzes a system
in which the intra-type triggering processes are all of the same efficiency while
all inter-type triggering processes have themselves the same efficiency
between themselves but in general weaker than the intra-type triggering processes.
We derive the time-dependence of the rates of events triggered from
a given shock for distributions of waiting times
of first-generation events  that have either exponential or power law tails.
Section 6 analyzes a system in which triggering between events
of different types proceeds through a one-dimension directed chain
of influence in type space.  We uncover a novel hierarchy of intermediate
asymptotic power law decays of the rate of triggered events as a function
of the distance of the events to the initial shock in the space of types.
Section 7 generalizes the results of section 6 by studying a
system in which triggering between events
of different types proceeds through a one-dimension symmetric chain
of influences in type space.  Two appendices give proofs and details
of the key results of the paper.

\section{Definitions and notations for the multivariate Hawkes processes}

\subsection{Monovariate Hawkes processes}

Self-excited conditional Poisson processes
generalize the cluster models by allowing each event, including
cluster members, i.e., aftershocks, to trigger their own events according to some memory
kernel $h(t-t_i)$.
\begin{equation}
\lambda(t | H_t, \Theta) = \lambda_c(t) + \sum_{i | t_{i} < t}  f(t-t_i)~,
\label{hyjuetg2tgj}
\end{equation}
where the history $H_t = \{ t_i \}_{1 \leq i \leq i_t,~ t_{i_t} \leq t < t_{i_t+1} }$ includes all events
that occurred before the present time $t$ and the sum
in expression (\ref{hyjuetg2tgj}) runs over all past triggered events.  The set of parameters is 
denoted by the symbol $\Theta$.
The term $\lambda_c(t)$ means
that there are some external background sources occurring according to a Poisson process
with intensity $\lambda_c(t)$, which may be a function of time, but all other events can be both triggered
by previous events and can themselves trigger their offsprings.  This gives rise
to the existence of many generations of events.

Introducing ``marks'' or characteristics for each event leads to a first multidimensional
extension of the self-excited process (\ref{hyjuetg2tgj}). The generalization consists in associating
with each event some marks (possible multiple traits), drawn from some distribution $p(m)$,
usually chosen invariant as a function of time:
\begin{equation}
\lambda(t, M| H_t, \Theta) = p(M) \left( \lambda_c(t) + \sum_{i | t_{i} < t}  f(t-t_i, M_i) \right)~,
\label{hyjuetgq2h56h2tgj}
\end{equation}
where the mark $M_i$ of a given previous event now controls the shape and properties
of the triggering kernel describing the future offsprings of that event $i$.
The history now consists in the set of occurrence times of each triggered event and their
marks: $H_t = \{ t_i, M_i\}_{1 \leq i \leq i_t,~ t_{i_t} \leq t < t_{i_t+1} }$. The first factor $p(M)$ in the r.h.s. of
expression (\ref{hyjuetgq2h56h2tgj}) writes that the marks of triggered events
are drawn from the distribution $p(M)$, independently of their generation and waiting times.
This is a simplifying specification, which can be relaxed. Inclusion of a spatial kernel
to describe how distance impacts triggering efficiency is straightforward.

A particularly well-studied specification of this class of marked self-excited point process
is the so-called Epidemic-Type-Aftershock-Sequence (ETAS) model \cite{KK81,Ogata88}:
\begin{equation}
\lambda(t, M | H_t, \Theta) = p(M) \left( \lambda_c + \sum_{i | t_{i} < t}   { k e^{a(M_i-M_0)} \over (t-t_i + c)^{1+\theta} }  \right)~,
\label{yj4ujbg}
\end{equation}
where $p(m)$ is given by the Gutenberg-Richter law, which
describes the probability density function (pdf)
of earthquakes of a given energy $E$, as being a power law, which translates
into an exponential function $p(M) \sim e^{-b M}$ of the log-energy scales inscribed in magnitudes $M$'s.
The memory kernel is often chosen as the power law (called the Omori law) with exponent $1+\theta$
and usually $0 \leq \theta <1$.
The time constant $c$ ensures finiteness of the triggering rate immediately
following any event.  Other forms with shorter memory, as the exponential, are also common.
Each event (of magnitude $M \geqslant M_0$) triggers
other events with a rate $\sim e^{a M}$, which defines the so-called fertility or productivity law. The lower magnitude
cut-off $M_0$ is such that events with marks smaller than $M_0$ do not generate offsprings.
This is necessary to make the theory convergent and well-defined, otherwise the crowd of small events may actually
dominate \cite{Helmsmallearth03,SorWerner05}. The constant $k$ controls the overall productivity law
and thus the average branching ratio defined by expression (\ref{eqpourn}) below.
The set of parameters is $\Theta = \{ b, \lambda_c, k, a, M_0, c, \theta \}$.

From a theoretical point of view,
the Hawkes models with marks has been studied in essentially two directions:
 (i) statistical estimations of its parameters with corresponding residual analysis as goodness of fits \cite{Ozaki79,Ogate81lewis,OgataAkaike82,Ogata83estmi,Zhuangetal02,ZhuangOgata04,OgaZhu06,Zhuangthesis,Marsan08,SorUtkindeclus};
 (ii) statistical properties of its space-time dynamics \cite{HS02,SaiSor07,SorHelmsing02,Helmsordiffth,Helmsorgrasso,Helmsordirect03,HelmsorPredict03,Saichevsor05,SaiHelmSor05,SaiSorlifetime,SaiSorrenoetas,SorUtkinSai08}.

The advantage of the self-excited conditional Hawkes process is to provide
a very parsimonious description
of the complex spatio-temporal organization of systems characterized
by self-excitatied bursts of events,
without the need to invoke ingredients other than the generally well-documented
stylized facts on the distribution of event sizes, the temporal ``Omori law''
for the waiting time before excitation of a new event and the productivity law controlling
the number of triggered events per initiator.

Self-excited models of point processes with additive structure of their intensity
on past events as in (\ref{hyjuetgq2h56h2tgj}) and (\ref{yj4ujbg}) \cite{Hawkes4} make them part
of the general family of branching processes \cite{Harris63}. The
crucial parameter is then the average branching ratio $n$, defined as the mean number of events of first
generation triggered per event. Using the notation of expression (\ref{yj4ujbg}), the average
branching ratio is given by
\begin{equation}
n = {k \over \theta c^\theta} \cdot {b \over b -a}~.
\label{eqpourn}
\end{equation}
Depending on applications, the branching ratio $n$ can vary with time,
from location to location and from type to type (as we shall see below
for the multivariate generalization). The branching ratio provides
a diagnostic of the susceptibility of the system to trigger activity in the
presence of some exogenous nucleating events.

Precise analytical results and numerical simulations show the existence of
three time-dependent regimes, depending on the ``branching ratio'' $n$ and on the sign of $\theta$.
This classification is valid for the range of parameters $a < b$.
When the productivity exponent $a$ is larger than the exponent $b$ of the Gutenberg-Richter law,
formula (\ref{eqpourn}) does not make sense anymore, which reflects the existence of
an explosive regime associated with stochastic finite-time singularities \cite{SorHelmsing02},
a regime that we do not consider further below, but which is relevant
to describe the accelerated damage processes leading to global systemic failures
in possibly many different types of systems \cite{SorPredictPNAS02}.
\begin{enumerate}
\item For $n<1$ (sub-critical regime), the rate of events triggered by a given shock
decays according to an effective Omori power law $\sim 1/t^p$, characterized by
a crossover from an Omori exponent $p=1-\theta$ for $t<t^*$ to a larger exponent
$p=1+\theta$ for $t>t^*$ \cite{HS02}, where $t^*$
is a characteristic time $t^* \simeq c / (1-n)^{1/\theta}$, which is controlled by the distance of $n$ to $1$.

\item For $n>1$ and $\theta>0$ (super-critical regime), one finds a transition from an
Omori decay law with exponent $p=1-\theta$ at early times since the mainshock
 to an explosive exponential increase of the activity rate at times
 $t> t^* \simeq c / (n-1)^{1/\theta}$ \cite{HS02,SaiSor10dissect}.

\item In the case $\theta<0$, there is a transition from an Omori law with exponent
$1-|\theta|$ similar to the local law, to an exponential increase at large times,
with a crossover time $\tau$ different from the characteristic
time $t^*$ found in the case $\theta>0$.
\end{enumerate}
We refer in particular to Ref.~\cite{SorOsorio10} for a short review of the main
results concerning the statistical properties of the space-time dynamics
of self-excited marked Hawkes conditional Poisson processes.

\subsection{Multivariate Hawkes processes}

The Multivariate Hawkes Process generalizes expressions (\ref{hyjuetgq2h56h2tgj}) and (\ref{yj4ujbg}) into
the following general form for the conditional Poisson intensity for an event of type $j$ among a set
of $m$ possible types (see the document \cite{Liniger09} for an extensive review):
\begin{equation}
\lambda_j(t | H_t) =  \lambda_j^0(t) + \sum_{k =1}^m  \Lambda_{kj} \int_{(-\infty, t) \times {\cal R}}
f_{k,j}(t-s) ~g_k(x)  ~N_k(ds \times dx)~,
\label{hyjuetgq2ujuk42tgj}
\end{equation}
where $H_t$ denotes the whole past history up to time $t$, $ \lambda_j^0$ is
the rate of spontaneous (exogenous) events of type $j$, i.e., the sources or immigrants of type $j$, $ \Lambda_{kj}$
is the $(k,j)$'s element of
the matrix of coupling between the different types which quantifies the ability of a type $k$-event
to trigger a type $j$-event. Specifically,
the value of an element $\Lambda_{kj}$ is just the average
number of first-generation events of type $j$ triggered by an event of type $k$.
This generalizes the branching ratio $n$ defined by (\ref{eqpourn}).
The memory kernel $f_{k,j}(t-s)$ gives the probability that an event of type $k$ that
occurred at time $s<t$ will trigger an event of type $j$ at time $t$.
The function $f_{k,j}(t-s)$ is nothing but the distribution of waiting times $t-s$ between the
impulse of event $k$ which impacted the system at some time $s$ and the occurrence
of an event of type $j$ at time $t$. The fertility (or productivity) law
$g_k(x)$ of events of type $k$ with mark $x$ quantifies the total average number of
first-generation events of any type triggered by an event of type $k$.
We have used the standard notation $ \int_{(-\infty, t) \times {\cal R}} f(t,x) N(ds \times dx) :=
\sum_{i | t_i < t} f(t_i, x_i)$.

The matrix ${\Lambda_{kj}}$ embodies both the topology of the network of interactions between
different types, and the coupling strength between
elements. In particular, ${\Lambda_{kj}}$  includes
the information contained on the adjacency matrix of the underlying network.
Analogous to the condition $n<1$ (subcritical regime) for the stability and stationarity of the monovariate Hawkes process,
the condition for the existence and stationarity of the process defined by (\ref{hyjuetgq2ujuk42tgj}) is that
the spectral radius of  the matrix ${\Lambda_{kj}}$ be less than $1$.
Recall that the spectral radius of a matrix is nothing but its largest eigenvalue.

To our knowledge, all existing works on the multivariate Hawkes processes assume that
the first moment of the memory kernel $f_{k,j}(t-s)$ exists. In the notations analogous to those
of equation (\ref{yj4ujbg}), if the memory kernels have an Omori like power law tail
$\sim 1/t^{1+\theta}$, this first-order moment condition imposes that
$\theta > 1$.  But, this is not the correct regime of parameters for earthquakes as well
as for other social epidemic processes, which have been shown to be characterized
by long-memory processes with $0 < \theta <1$ \cite{Amazonbook1,Amazonbook2,Sorendoexo05,Cranesor08}.
This regime $0 < \theta <1$ leading to infinite first-order moments leads to very rich
new scaling behaviors in the multivariable case,
as we are going to show below. Actually, we will derive
the remarkable results that multivariate Hawkes processes can be characterized
by a hierarchy of dynamics with different exponents, all related to the fundamental
Omori law for first generation waiting times.

\section{Temporal multivariate generating moment function (GMF)}

\subsection{Generating moment function for the cumulative number of first-generation events triggered until time $t$}

Among the $m$ types of events, consider the $k$-th type and its first generation offsprings.
Let us denote $R_1^{k,1}(t) , R_1^{k,2}(t) ,\dots, R_1^{k,m}(t)$, the cumulative number of ``daughter'' events
of first generation of type $1, 2, \dots, m$ generated by this ``mother'' event of type $k$ from time $0$ until time $t$.
With these notations, the generating moment function (GMF) of all events of first generation
that are triggered by a mother event of type $k$ until time $t$ reads
\begin{equation}\label{4a}
A_1^k(y_1,y_2,\dots,y_m;t) := \text{E}\left[ \prod_{s=1}^m y_s^{R_1^{k,s}(t)}\right] ~,
\end{equation}
where $\text{E}\left[ X \right]$ denotes the average of $X$ over all possible  statistical realizations.
We shall also need the definition of the GMF $A_1^k(y_1,y_2,\dots,y_m)$
defined by
\begin{equation}
A_1^k(y_1,y_2,\dots,y_m) := \text{E}\left[ \prod_{s=1}^m y_s^{R_1^{k,s}}\right] ~,
\label{yjrukenw}
\end{equation}
where $R_1^{k,1}= {\rm lim}_{t \to +\infty}~R_1^{k,1}(t) , R_1^{k,2}={\rm lim}_{t \to +\infty}~R_1^{k,2}(t),\dots, R_1^{k,m}={\rm lim}_{t \to +\infty}~R_1^{k,m}(t)$ are the cumulative number of ``daughter'' events
of first generation of type $1, 2, \dots, m$ generated by the ``mother'' event of type $k$ over all times.
One may rewrite this function in probabilistic form
\begin{equation}\label{1}
A_1^k(y_1,y_2,\dots,y_m) := \sum_{r_1=0}^\infty \dots \sum_{r_m=0}^\infty P_k(r_1,\dots,r_m) \prod_{s=1}^m y_s^{r_s}~,
\end{equation}
where $P_k(r_1,\dots,r_m)$ is the probability that the mother event of type $k$
generates $R^{k,1}=r_1$ first-generation events of type $1$, $R^{k,2}=r_2$ first-generation events of type $2$,
and so on.

The events are assumed to occur after waiting times between the mother event and their occurrences that are
mutually statistically independent and characterized by the probability density functions (pdf)
$\{f_{k,s}(t)\}$, where all $f_{k,s}(t)\equiv 0$ for $t<0$.
Let us denote $\mathcal{P}_k (d_1,d_2,\dots,d_m; t)$ the probability that
the cumulative numbers $\{R_1^{k,s}(t)\}$ up to time $t$ of first-generation events that have been
triggered by the mother event of type $k$ are equal to
\begin{equation}
R_1^{k,1}(t) = d_1 ~, \qquad R_1^{k,2}(t) = d_2~ , \quad \dots \quad R_1^{k,m}(t) = d_m~ .
\end{equation}
Let us relate this probability $\mathcal{P}_k (d_1,d_2,\dots,d_m; t)$
to that, denoted $\mathcal{P}_k (d_1,d_2,\dots,d_m;t|r_1,r_2,\dots,r_m)$, obtained
under the additional condition that the total numbers
of first-generation events that have been triggered by the mother event of type $k$ over the
whole time interval $t \to \infty$ are fixed at the values $\{r_{1},r_{2},\dots,r_{m}\}$. Obviously,
$\mathcal{P}_k (d_1,d_2,\dots,d_m;t|r_1,r_2,\dots,r_m)$ is given by a product of binomial distributions
\begin{equation}
\label{5}
\begin{array}{c} \displaystyle
\mathcal{P}_k (d_1,d_2,\dots,d_m;t|r_1,r_2,\dots,r_m) =
\prod_{s=1}^m \binom{r_s}{d_s} \mu_{k,s}^{d_s}(t) [1- \mu_{k,s}(t)]^{r_s-d_s} ~,
\\[4mm] \displaystyle
0\leqslant d_1 \leqslant r_1 , \qquad \dots \qquad 0\leqslant d_m \leqslant r_m~ ,
\end{array}
\end{equation}
where
\begin{equation}
\mu_{k,s}(t) = \int_0^t f_{k,s}(t')dt' ~.
\label{jk57ikoin}
\end{equation}

Knowing the conditional probabilities
$\mathcal{P}_k (d_1,d_2,\dots,d_m; t|r_1,r_2,\dots,r_m)$ \eqref{5},
one can calculate their unconditional counterparts using the following relation
\begin{equation}
\label{jhweqfbgq}
\begin{array}{c}\displaystyle
\mathcal{P}_k (d_1,d_2,\dots,d_m; t) =
\\[1mm]\displaystyle
\sum_{r_1=d_1}^\infty \dots \sum_{r_m=d_m}^\infty \mathcal{P}_k (d_1,d_2,\dots,d_m; t|r_1,r_2,\dots,r_m) P_k(r_1,r_2,\dots,r_m)  ~,
\end{array}
\end{equation}
where $P_k(r_1,r_2,\dots,r_m)$ is the probability that the total numbers
of first-generation events that have been triggered by the mother event of type $k$ over the
whole time interval $t \to \infty$ take the values $\{r_{1},r_{2},\dots,r_{m}\}$.

Substituting the relations \eqref{5} in (\ref{jhweqfbgq}) yields
\begin{equation}
\label{6a}
\begin{array}{c}\displaystyle
\mathcal{P}_k (d_1,d_2,\dots,d_m; t) =
\\[1mm]\displaystyle
\sum_{r_1=d_1}^\infty \dots \sum_{r_m=d_m}^\infty P_k(r_1,\dots,r_m)
\prod_{s=1}^m \binom{r_s}{d_s} \mu_{k,s}^{d_s}(t) [1- \mu_{k,s}(t)]^{r_s-d_s}~ .
\end{array}
\end{equation}
The interest in this expression (\ref{6a}) is that the GMF $A_1^k(y_1,y_2,\dots,y_m;t)$
defined by expression (\ref{4a}) can be rewritten in probabilistic form as
\begin{equation}
A_1^k(y_1,y_2,\dots,y_m;t) = \sum_{d_1=0}^\infty \dots \sum_{d_m=0}^\infty \mathcal{P}_k(d_1,\dots,d_m;t) \prod_{s=1}^m y_s^{d_s}~.
\label{thjkuikujwthq}
\end{equation}

We are now prepared to state the following theorem, which is essential for our subsequent derivations.
\begin{theorem}
\label{httjh6uju6}
The GMF $A_1^k(y_1,y_2,\dots,y_m;t)$ defined by expression (\ref{4a}) can be represented in the form
\begin{equation}
\label{6}
A_1^k(y_1,y_2,\dots,y_m;t) = Q_k[\mu_{k,1}(t)(y_1-1),\dots, \mu_{k,m}(t)(y_m-1)]~ ,
\end{equation}
where
\begin{equation}
\label{7}
Q_k(z_1,\dots,z_m):=A_1^k(1+z_1,\dots,1+z_s) ~~ \Leftrightarrow ~~
A_1^k(y_1,\dots,y_m) = Q_k(y_1-1,\dots,y_m-1)~ .
\end{equation}
\end{theorem}
The proof is given in Appendix A.

\subsection{GMF for the cumulative numbers of events over all generation triggered until time $t$}

Let us define the GMF
\begin{equation}\label{8}
A^k(y_1,y_2,\dots,y_m;t) := \text{E}\left[ \prod_{s=1}^m y_s^{R^{k,s}(t)}\right]~ ,
\end{equation}
where $\{R^{k,s}(t)\}$ is the total number of events summed over all generations of events of type $s$ triggered
by a mother event of type $k$ starting at time $0$ up to time $t$.

Due to the branching nature of the process,
the equation determining the GMF $\{A^k(y_1,y_2,\dots,y_m;t)\}$
is obtained by
\begin{enumerate}
\item replacing in the left-hand-side of expression \eqref{6} $A_1^k(y_1,y_2,\dots,y_m;t)$ by $A^k(y_1,y_2,\dots,y_m;t)$; this means that we deal with events of all generations occurring till current time $t$;

\item replacing in the right-hand-side of expression \eqref{6} the arguments $y_s$ by
\begin{equation}
y_s \quad \Rightarrow \quad y_s \int_0^t f_{k,s}(t'|t) A^s(y_1,\dots,y_m;t-t') dt' ~,
\label{hwjyjuki}
\end{equation}
where $\{f_{k,s}(t'|t)\}$ is the conditional pdf of the random times $\{t'_{k,s}\}$ of occurrence of some
first-generation event of type $s$ triggered by the mother event of type $k$, under the condition that
it occurred within the time interval $t'\in(0,t)$. The conditional pdf $\{f_{k,s}(t'|t)\}$ is given by
\begin{equation}
f_{k,s}(t'|t) ={ f_{k,s}(t')\over \mu_{k,s}(t)}~ ,
\end{equation}
where $\mu_{k,s}(t)$ is defined by (\ref{jk57ikoin}).
The pdf $f_{k,s}(t'|t)$ inside the integral (\ref{hwjyjuki})
takes into account that first-generation events  are occurring at random times
$t' < t$. The other factor $A^s(y_1,\dots,y_m;t-t')$
takes into account all-generation events that are triggered by some first-generation event from its appearance time $t'$ till the current time $t$.
\end{enumerate}

The equation for the GMF $\{A^k(y_1,y_2,\dots,y_m;t)\}$ is thus
\begin{equation}\label{9}
A^k(y_1,y_2,\dots,y_m;t) =
Q_k\left[B^{k,1}(y_1,\dots,y_m;t),\dots,B^{k,m}(y_1,\dots,y_m;t)\right] ~,
\end{equation}
where
\begin{equation}\label{10}
B^{k,s}(y_1,\dots,y_m;t) = \int_0^t f_{k,s}(t-t')[y_s A^s(y_1,\dots,y_m;t') -1]dt' ~.
\end{equation}

\section{General relations for the mean numbers of events over all generation triggered up to time $t$}

The set of equations \eqref{9} together with the relations \eqref{10} provides the basis for a full description
of the statistical and temporal properties of multivariate branching (Hawkes) processes.
Here, we restrict our attention to the average activities, by studying the temporal
dependence of the average number of events following the occurrence of a mother event of a given type.

The mean number of events of type $s$ over all generations counted until time $t$
that are triggered by a mother event of type $k$ that occurred at time $t=0$, defined by
\begin{equation}
\bar{R}^{k,s}(t):= \text{E}[R^{k,s}(t)]~,
\end{equation}
is  given by the relation
\begin{equation}\label{14}
\bar{R}^{k,s}(t)= {\partial\over\partial y_s} A^k(y_1,y_2,\dots,y_m;t)\big|_{y_1=\dots =y_m=1}~ ,
\end{equation}
where $A^k(y_1,y_2,\dots,y_m;t)$ satisfies to the set of equations \eqref{9}.

In order to derive the equations determining the set $\{\bar{R}^{k,s}(t)\}$,
we need to state the following properties exhibited by the functions $Q_k(y_1,\dots,u_m)$ given by \eqref{7} and
$B^{k,s}(y_1,y_2,\dots,y_n;t)$ given by \eqref{10}, which are contributing to equation \eqref{9}.
From the definition of the functions $B^{k,s}(y_1,y_2,\dots,y_n;t)$ and of the GMF $A^k(y_1,y_2,\dots,y_m;t)$, we have
\begin{equation}
B^{k,s}(y_1,y_2,\dots,y_n;t)\big|_{y_1=\dots =y_m=1} \equiv 0
\label{thjeynbqbq}
\end{equation}
and
\begin{equation}
{\partial\over\partial y_s} B^{k,\ell}(y_1,y_2,\dots,y_n;t,\tau)\big|_{y_1=\dots =y_m=1} = \mu_{k,s}(t) \cdot \delta_{\ell,s} + f_{k,\ell}(t)\otimes \bar{R}^{\ell,s}(t) ~.
\end{equation}
The convolution operation is defined as usual by $f(t) \otimes g(t) := \int_0^t f(t-t') g(t') dt'$.
Moreover, the following equality holds
\begin{equation}
{\partial\over\partial y_s} Q_k (y_1,y_2,\dots,y_n;t,\tau)\big|_{y_1=\dots =y_m=0} = n_{k,s}~ ,
\label{ytjujsw2gj}
\end{equation}
where $n_{k,s}$ is the mean value of the total number of first-generation events of type $s$
triggered by a mother of type $k$. The set $\{n_{k,s}\}$ for all $k$'s and $s$'s generalize the average
branching ratio $n$ defined by expression (\ref{eqpourn}) above for monovariate branching processes and are given by
\begin{equation}\label{2}
n_{k,s} = {\partial\over\partial y_s} A_1^k(y_1,y_2,\dots,y_m)\big|_{y_1=\dots =y_m=1}
\end{equation}
where $A_1^k(y_1,y_2,\dots,y_m)$ is defined in (\ref{yjrukenw}).

Using the above relations (\ref{thjeynbqbq}-\ref{ytjujsw2gj}),
the equality \eqref{14} together with equation \eqref{9} yields
\begin{equation}\label{15}
\bar{R}^{k,s}(t)= n_{k,s}\cdot \mu_{k,s}(t) + \sum_{\ell=1}^m n_{k,\ell} f_{k,\ell}(t)\otimes \bar{R}^{\ell,s}(t) ~.
\end{equation}
Introducing the event rates, i.e., the number of events per unit time,
\begin{equation}\label{ratesdef}
\rho^{k,s}(t) = {d\bar{R}^{k,s}(t)\over dt} ~,
\end{equation}
expression \eqref{15} transforms into
\begin{equation}\label{rhokseqs}
\rho^{k,s}(t)= n_{k,s}(t) + \sum_{\ell=1}^m n_{k,\ell}(t)\otimes \rho^{\ell,s}(t)~ ,
\end{equation}
where we have used the following notation
\begin{equation}
n_{k,s}(t):= n_{k,s}\cdot f_{k,s}(t)~ .
\end{equation}
The set of equations (\ref{rhokseqs}) for all $k$'s and $s$'s constitute the fundamental starting
point of our analysis.

In order to make further progress, in view of the convolution operator, it is convenient to
work with the Laplace transform of the event rates:
\begin{equation}\label{laptransfdef}
\tilde{\rho}^{k,s}(u) = \int_0^\infty \rho^{k,s} (t) e^{-ut} dt ~.
\end{equation}
Introducing the matrices
\begin{equation}\label{phinmatrix}
\tilde\Phi(u) = [\tilde{\rho}^{k,s}(u)]  \quad \text{and} \quad \tilde{N}(u) = [\tilde{n}_{k,s}(u)] ~,
\end{equation}
we obtain the following equation for the matrix $\tilde\Phi(u)$
\begin{equation}
\hat{I} ~ \tilde{\Phi}(u) = \tilde{N}(u) + \tilde{N}(u) ~ \tilde{\Phi}(u) ~,
\end{equation}
whose solution is
\begin{equation}\label{rhomatrixsol}
\tilde{\Phi}(u) = {\tilde{N}(u) \over \hat{I} - \tilde{N}(u)}~ .
\end{equation}
The rest of the paper is devoted to the study of this solution (\ref{rhomatrixsol}) for various system structures
and memory kernels.

\section{Symmetric mutual excitations}

\subsection{Definitions \label{tjujs}}

Let us consider the case where the set $\{n_{k,s}\}$ defined by expression (\ref{2}) reduces to
\begin{equation}
n_{k,k} = a ~; \qquad  \qquad n_{k,s} = b , ~~ k\neq s ~,
\label{kumnw}
\end{equation}
This form (\ref{kumnw}) means that events of a given type have identical triggering efficiencies quantified by $a$
to generate first-generation events of the same type.  They also have identical efficiencies quantified by $b$
to trigger first-generation events of a different type.
As a consequence, the mean number of first-generation events of all kinds that are triggered by
a mother event of some type $k$,
\begin{equation}
n_k = \sum_{s=1}^m n_{k,s}~ ,
\label{tjetmnw}
\end{equation}
is independent of $k$ and given by
\begin{equation}\label{ambn}
n_k  =  n = a + (m-1) b ~, ~~~~{\rm for~all}~k~.
\end{equation}
Introducing the factor
\begin{equation}\label{qba}
q = {b\over a}
\end{equation}
comparing the inter-types with the intra-type triggering efficiencies, we obtain
\begin{equation}\label{abdefs}
a = {n\over 1+(m-1) q} ~, \qquad b = {n q \over 1+(m-1) q}~ .
\end{equation}

In the time domain, we consider the case of symmetric mutual excitations such that all pdf's $f_{k,s}(t)\equiv f(t)$
are independent of the indexes $k$ and $s$ and are all equal to each other.

\subsection{General solution in terms of Laplace transforms}

With the definitions of subsection \ref{tjujs}, it follows that $\tilde{n}_{k,s}(u) = n ~\tilde{f}(u)$
and one can show that all diagonal and non-diagonal entries of the matrix $\tilde\Phi(u)$
given by \eqref{rhomatrixsol} are given respectively by
\begin{equation}\label{rhomatrudiag}
\begin{array}{c} \displaystyle
\tilde{\rho}(u):=\tilde{\rho}^{k,k}(u) = {n\tilde{f}(u)\over 1-n\tilde{f}(u)} \cdot {1+ n\tilde{f}(u) (q-1)\over 1+n\tilde{f}(u)(q-1) + q (m-1)}~ ,
\\[4mm] \displaystyle
\tilde{g}(u):=\tilde{\rho}^{k,s}(u) =  {n\tilde{f}(u)\over 1-n\tilde{f}(u)} \cdot {q \over 1+n\tilde{f}(u)(q-1) + q (m-1)}~, ~~~~k\neq s~ .
\end{array}
\end{equation}
Moreover, the Laplace transform
\begin{equation}
\tilde{\rho}^k(u) = \sum_{s=1}^m \tilde{\rho}^{k,s}(u) =\tilde{\rho}(u) +(m-1) \tilde{g}(u)
\end{equation}
of the total rate of events of all types triggered by a mother jump of type $k$ defined by $\rho^{k}(t) =\sum_{s=1}^m \rho^{k,s}(t)$
satisfies the relation
\begin{equation}
\tilde{\rho}^k(u) =  {n\tilde{f}(u)\over 1-n\tilde{f}(u)}~ .
\label{heytnb}
\end{equation}

\subsection{Exponential pdf of triggering times of first-generation events}

Let us first study the case where the pdf $f(t)$ of the waiting times to generate
first-generation events is exponential:
\begin{equation}\label{ftexp}
f(t) = \alpha e^{-\alpha t}  \qquad \iff \qquad \tilde{f}(u) = {\alpha\over \alpha+ u} .
\end{equation}
The inverse Laplace transforms of the solutions \eqref{rhomatrudiag} are then
\begin{equation}\label{rhokktexpr}
\begin{array}{l} \displaystyle
\rho(t) := \rho^{k,k}(t) = \alpha {n\over m}e^{-(1-n) \alpha t}\left(1 +
(m-1) \gamma e^{- n (1-\gamma) \alpha t}\right)~ ,
\\[4mm]\displaystyle
g(t) := \rho^{k,s}(t) = \alpha {n\over m}e^{-(1-n) \alpha t} \left(1-
\gamma e^{- n (1-\gamma) \alpha t}\right) , \qquad k\neq s~ ,
\end{array}
\end{equation}
where
\begin{equation}
\gamma:=\gamma(q,m) = {1-q\over 1+(m-1) q} .
\end{equation}
Expressions (\ref{rhokktexpr}) give the explicit time dependence of two functions $\rho(t)$ and $g(t)$:
\begin{itemize}
\item $\rho(t):= \rho^{k,k}(t)$ is the rate of events over all generations of some type $k$ resulting from
a given mother event of the same type $k$. Notice that the term ``over all generations'' means that
an event of type $k$ occurring at some time $t>0$, and belonging to the descent
of some previous mother of the same type $k$ that occurred at time $0$,
may have been generated through a long cascade
of intermediate events of possibly different kinds, via a kind of inter-breeding genealogy.
\item $g(t):= \rho^{k,s}(t)$ is the rate of events over all generations of some type $s$ resulting from
a given mother event of a different type $k$. As for $\rho(t)$,
an event of type $s$ occurring at some time $t>0$, and belonging to the descent
of some previous mother of a different type $k$ that occurred at time $0$,
may have been generated through a long cascade
of intermediate events of possibly different kinds, via a kind of inter-breeding genealogy.
\end{itemize}

Figures~1 and 2 show respectively $\rho(t):= \rho^{k,k}(t)$ and $g(t):= \rho^{k,s}(t)$
for the case of $m=3$ types of events and rather close to criticality ($n=0.99$), for different
coupling amplitudes $q=1; 0.1;0.01$.

\begin{quote}
\centerline{
\includegraphics[width=11cm]{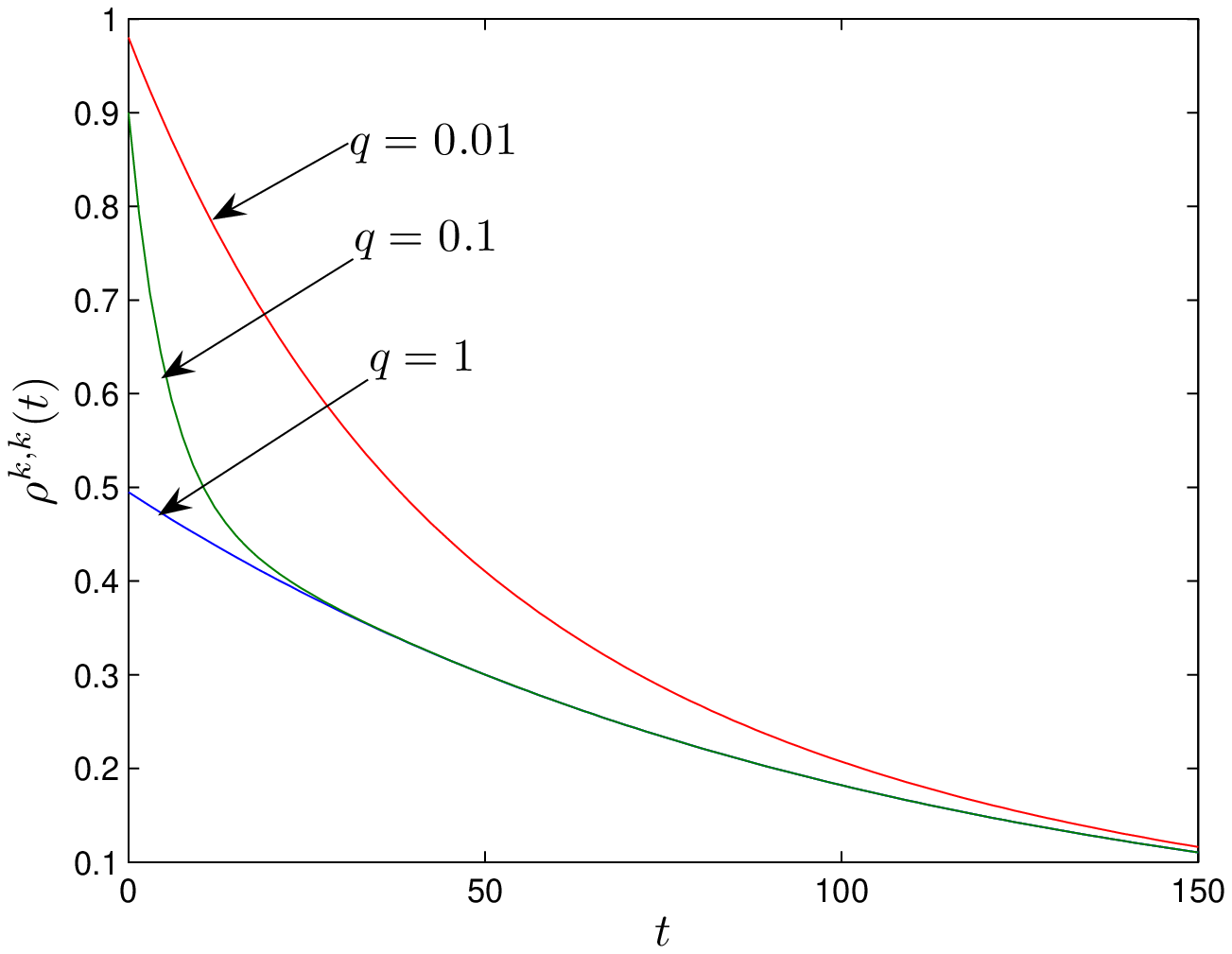}}
{\bf Fig.~1:} \small{Time dependence of $\rho(t):= \rho^{k,k}(t)$, for $m=3$, $n=0.99$, and $q=1; 0.1;0.01$. Time
is in unit of $1/\alpha$.}
\end{quote}

\begin{quote}
\centerline{
\includegraphics[width=11cm]{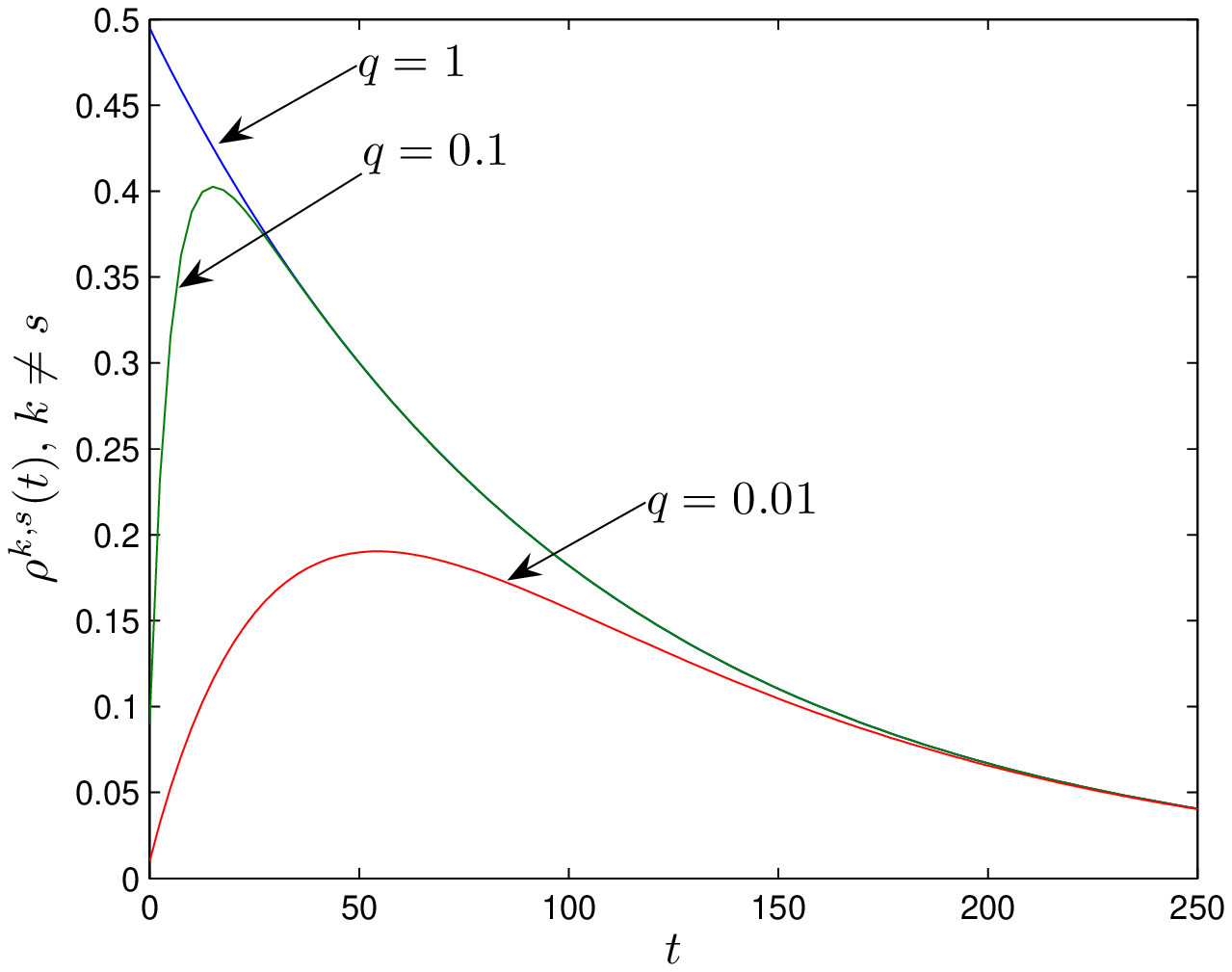}}
{\bf Fig.~2:} \small{Time dependence of $g(t):= \rho^{k,s}(t)$, for $m=3$, $n=0.99$ and $q=1; 0.1;0.01$. Time
is in unit of $1/\alpha$.}
\end{quote}

The case $q=1$ of complete coupling is special in two ways: (i) $\rho(t)$ and $g(t)$ are proportional to each other; (ii)
there is only one time scale $\tau_1 ={1 \over \alpha} \cdot {1 \over 1-n}$. In contrast, as soon as $q<1$, i.e., events of a given type
tends to trigger more events of the same type than events of different types, one can observe that the
time dependence of $\rho(t)$ and $g(t)$ become qualitatively different. The monotonous decay of $\rho(t)$ can be
contrasted with the non-monotonous bell-shape dependence of $g(t)$. This non-monotonous behavior
of $g(t)$ results from the progressive seeding of events of a different type than the initial mother type
by the less efficient mutual excitation process. This is associated with the introduction of a second time
scale $\tau_2 = {1 \over \alpha} \cdot {1 \over 1-n\gamma}  \leq \tau_1$ controlling the dynamics
of both $\rho(t)$ and $g(t)$ at short times. This effect is all the stronger, the smaller $q$ is, i.e., the larger $\gamma$ is.

This phenomenon of the occurrence of a second time scale $\tau_2$ and of the distinct behavior
of  $g(t):= \rho^{k,s}(t)$ for $q<1$ constitutes a characteristic signature of the mutual excitation mechanism,
in a system in which the time response function is exponential (Poisson, i.e., without memory).
The cascade of triggering together with the inter-breeding of events of different types break the Poissonian
nature of the relaxation of the event activity triggered by a given mother ancestor. This is different
from what occurs for a single event type $m=1$ for which the existence of multiple generations
do not change the Poissonian nature of the relaxation process. It only extend the time scale
according to $\tau_1 ={1 \over \alpha} \cdot {1 \over 1-n}$ as the average branching ratio $n$ increases to the
critical value $1$.

\subsection{Power law pdf of triggering times of first-generation events}

The same qualitative picture emerges for other pdf's such as  power laws, with a monotonous decay of $\rho(t)$ coexisting with
a growth from zero up to a maximum followed by a decay for $g(t)$. But more interesting features appear,
such as the renormalization of the exponents in two distinct families, as we now show. As many systems exhibit
power law pdf's of waiting times with rather small exponents $\theta \leq 0.5$ (see definition in
expression (\ref{fpffpowerdef})), we shall consider this regime
in the following.

For definiteness, we consider the pdf with power law tail given by
\begin{equation}\label{fpffpowerdef}
f(t) = {\alpha\theta \over (1+\alpha t)^{1+\theta}} , \qquad t> 0~, \qquad  \theta >0~.
\end{equation}
The constant $1$ in the denominator regularizes the pdf at times $t < 1/\alpha$.
The corresponding Laplace transform of (\ref{fpffpowerdef}) is
\begin{equation}\label{tildafpowlaw}
\tilde{f}(u) = \theta~ e^{u/\alpha} \left({u\over\alpha}\right)^\theta \Gamma\left(-\theta, {u\over\alpha}\right) ~,
\end{equation}
which can then be used in (\ref{rhomatrudiag}) to get the general solutions.

\subsubsection{Non-critical same-type activity rate $\rho(t)=:\rho^{k,k}(t)$}

We first rewrite the Laplace transform $\tilde{\rho}(u)$ given by \eqref{rhomatrudiag} in a form more convenient for its analysis:
\begin{equation}\label{tilrhothruphi}
\tilde{\rho}(u) = \bar{R}^{k,k} \cdot {1-\varphi(u)\over 1+\gamma_0 \varphi(u)}
\cdot {1+\gamma_1 \varphi(u)\over 1+\gamma_2 \varphi(u)}~ ,
\end{equation}
where
\begin{equation}\label{varphidef}
\varphi(u) := 1-\tilde{f}(u)
\end{equation}
and
\begin{equation}\label{gammasdef}
\begin{array}{c} \displaystyle
\gamma_0 = {n \over 1-n} , \quad \gamma_1 = {n (1-q)\over 1-n+ q n} , \quad \gamma_2 = {n(1-q)\over 1-n + q (n+m-1)} ,
\\[5mm] \displaystyle
\gamma_0> \gamma_1 > \gamma_2 \qquad (q>0,~ m>1) .
\end{array}
\end{equation}

The long time behavior of $\rho(t)$ is controlled by the small $u$ properties of $\tilde{\rho}(u)$, itself dependent
on the behavior of $\varphi(u)$ for small $u$. From its definition (\ref{varphidef}), we have
that $\varphi(u) \to 0$ for $u \to 0$. Then, the asymptotic behavior of the Laplace transform \eqref{tilrhothruphi} of $\rho(t)$ is
\begin{equation}\label{tilrhoutozero}
\tilde{\rho}(u) \sim \bar{R}^{k,k} - \bar{R}^{k,k} \cdot \gamma_\rho \varphi(u)~,
\qquad \gamma_\rho = 1+\gamma_0 + \gamma_2 -\gamma_1~ .
\end{equation}

For $\theta\in(0,1)$, the auxiliary function $\varphi(u)$ \eqref{varphidef} has the following asymptotic behavior
\begin{equation}\label{varphiasymp}
\varphi(u) \sim \beta v^\theta\ll 1 ~, \qquad v\ll 1~, \qquad \beta=\Gamma(1-\theta)~ , \quad v = {u\over\alpha}~ .
\end{equation}
Accordingly, relation \eqref{tilrhoutozero} transforms into
\begin{equation}\label{tilrhouinf}
\tilde{\rho}(u) \sim \bar{R}^{k,k} \left[1-   \gamma_\rho \beta v^\theta \right]~ .
\end{equation}
The corresponding asymptotic of the rate $\rho(t)$ is thus
\begin{equation}\label{ratettoinfty}
\rho(t) \sim \bar{R}^{k,k} \cdot{\gamma_\rho\beta \theta \over \Gamma(1-\theta)} ~
{1 \over \left(\alpha t\right)^{1+\theta}}= \bar{R}^{k,k} \cdot \gamma_\rho \theta
{1 \over \left(\alpha t\right)^{1+\theta}} ~, \qquad t\to\infty~ .
\end{equation}
The rate $\rho(t):=\rho^{k,k}(t)$ of events over all generations of some type $k$ resulting from
a given mother event of the same type $k$ that occurred at time $0$ decays with the same power law behavior
as the bare memory function or pdf of waiting times for first-generation events.
The only significant difference is the renormalization of the amplitude by the factor
$\bar{R}^{k,k} \cdot \gamma_\rho$ resulting from the cascades of generations and inter-breeding
between the different event types.

\subsubsection{Intermediate critical asymptotic same-type activity rate $\rho(t):=\rho^{k,k}(t)$ \label{tjkiik}}

For $n$ close to $1$, $\gamma_0$ becomes large and there is an interesting intermediate asymptotic regime
describing the intermediate time decay of $\rho(t)$.  To describe it, we need to distinguish the following three
parameter regimes.
\begin{enumerate}
\item $\gamma_0 \gg 1$ and $\gamma_2 < \gamma_1 \simeq 1$. This occurs for $n \to 1$ while $q$ is not too
close to $0$. For instance, $n=0.95$, $q=0.5$, $m=5$ yield $\gamma_2 = 0.613, \gamma_1=0.905, \gamma_0=19$.
In this case, there is an intermediate range on the $u$-axis defined by
\begin{equation}\label{intermediateu1}
\left(\gamma_0 \beta\right)^{-\theta} \ll v \ll \left(\gamma_1 \beta\right)^{-\theta}
\quad \iff \quad \alpha \left(\gamma_0 \beta\right)^{-\theta} \ll u \ll \alpha \left(\gamma_1 \beta\right)^{-\theta}  \approx \alpha~,
\end{equation}
such that $\gamma_0 \varphi(u) \gg 1$ while
$\gamma_1 \varphi(u) \ll 1$ and $\gamma_2 \varphi(u) \ll 1$. In this range, the leading
terms controlling the value of expression (\ref{tilrhothruphi}) is
\begin{equation}\label{tilrhothwreg2ruphi}
\tilde{\rho}(u) \approx  \bar{R}^{k,k} \cdot {1\over \gamma_0 \varphi(u)} ~.
\end{equation}
Substituting the asymptotic relation \eqref{varphiasymp}, we obtain
\begin{equation}
\tilde{\rho}(u) \approx  \bar{R}^{k,k} \cdot {1 \over \gamma_0 \Gamma(1-\theta)} ~v^{-\theta}~.
\end{equation}
The corresponding intermediate asymptotic of the rate is
\begin{equation}\label{rateintasymp1}
\rho(t) \approx \bar{R}^{k,k} {1 \over \gamma_0} {\sin(\pi\theta)\over\pi} {1 \over (\alpha t)^{1-\theta}}~ ,
\qquad 1 \ll \alpha t \ll \left[\gamma_0 \Gamma(1-\theta)\right]^{1/\theta} ~.
\end{equation}

\item $\gamma_0 > \gamma_1 \gg 1$ and $\gamma_2 \simeq 1$: This occurs for $n \to 1$ with $q$ close to $0$
and $m$ large. For instance, $n=0.95$, $q=0.01$, $m=50$ yield $\gamma_2 = 1.68, \gamma_1=15.8, \gamma_0=19$.
In this case, there is an intermediate range on the $u$-axis defined by
\begin{equation}\label{intermediateu2}
\left(\gamma_1 \beta\right)^{-\theta} \ll v \ll \left(\gamma_2 \beta\right)^{-\theta}
\quad \iff \quad \alpha \left(\gamma_1 \beta\right)^{-\theta} \ll u \ll  \alpha \left(\gamma_2 \beta\right)^{-\theta} \approx \alpha ~,
\end{equation}
such that $\gamma_0 \varphi(u) > \gamma_1 \varphi(u) \gg 1$ while
$\gamma_2 \varphi(u) \ll 1$. In this range, the leading
terms controlling the value of expression (\ref{tilrhothruphi}) is
\begin{equation}
\tilde{\rho}(u) = \bar{R}^{k,k} \cdot {\gamma_1 \over \gamma_0} \cdot (1 -  \varphi(u))~ ,
\end{equation}
whose inverse Laplace transform has the same form as (\ref{ratettoinfty}) with $\gamma_\rho$ replaced by $1$.

\item $\gamma_0 > \gamma_1 > \gamma_2  \gg 1$: This occurs for $n \to 1$ with $q$ close to $0$
and $m$ not too large. For instance, $n=0.95$, $q=0.01$, $m=2$ yield $\gamma_2 = 13.5, \gamma_1=15.8, \gamma_0=19$.
Then, in the intermediate interval on the $u$-axis
\begin{equation}\label{intermediateu3}
\left(\gamma_2 \beta\right)^{-\theta} \ll v \ll 1 \quad \iff \quad \alpha \left(\gamma_2 \beta\right)^{-\theta} \ll u \ll \alpha ~,
\end{equation}
the asymptotic relation \eqref{varphiasymp} holds, while at the same time $\gamma_2 \varphi(u)\gg 1$.
In this range (\ref{intermediateu3}), the leading
terms controlling the value of expression (\ref{tilrhothruphi}) is
\begin{equation}\label{tilrhointermedrel}
\tilde{\rho}(u) \approx \bar{R}^{k,k} \cdot {1\over \gamma_0 \varphi(u)} \cdot {\gamma_1 \varphi(u)\over \gamma_2 \varphi(u)} = \bar{R}^{k,k} \cdot {\gamma_1\over \gamma_0 \gamma_2} \cdot {1\over \varphi(u)}~ .
\end{equation}
Using the asymptotic relation \eqref{varphiasymp}, we obtain
\begin{equation}
\tilde{\rho}(u) \approx \mathcal{G}~ v^{-\theta} , \qquad \mathcal{G} = \bar{R}^{k,k} \cdot {\gamma_1\over \gamma_0 \gamma_2 \beta} = \bar{R}^{k,k} \cdot {\gamma_1\over \gamma_0 \gamma_2 \Gamma(1-\theta)}~.
\end{equation}
The corresponding intermediate power asymptotic of the rate $\rho(t)$ is
\begin{equation}\label{rateintasymp3}
\rho(t) \approx \bar{R}^{k,k} {\gamma_1\over \gamma_0\gamma_2} {\sin(\pi\theta)\over\pi}
{1 \over (\alpha t)^{1-\theta}}~ , \qquad 1 \ll \alpha t \ll \left[\gamma_2 \Gamma(1-\theta)\right]^{1/\theta} ~.
\end{equation}
As an illustration, for the above values $n=0.95$, $q=0.01$, $m=2$, the power law (\ref{rateintasymp3})
with exponent $1-\theta$ holds up to a maximum time
$\left[\gamma_2 \Gamma(1-\theta)\right]^{1/\theta} \alpha^{-1} \approx 75,000  \alpha^{-1}$ for $\theta=0.25$~.
\end{enumerate}

Let us summarize and interpret the above results,
\begin{enumerate}
\item $n \to 1$ and $q$ not small ($\gamma_0 \gg 1$ and $\gamma_2 < \gamma_1 \simeq 1$).
The intermediate power law asymptotic (\ref{rateintasymp1}) with exponent $1-\theta$ is similar
to the renormalized response function due to the cascade of generations found for the self-excited
Hawkes process with just one type of events \cite{SS99,HS02,SaiSor10dissect}. The mechanism is the same,
since a coupling coefficient $q$ not too small ensures a good mixing among all generations.

\item $n \to 1$ with $q \to 0$ and $m \to \infty$ with $qm \simeq 1$ ($\gamma_0 > \gamma_1 \gg 1$ and $\gamma_2 \simeq 1$).
In contrast with the previous case, the activity rate $\rho(t)$ exhibits the same decay (\ref{ratettoinfty}) with exponent $1+\theta$
as if the system was far from criticality. In a sense, due to the weak mutual triggering efficiency and the many event types,
the system is never critical.

\item $n \to 1$ with $q \to 0$ and $m$ not too large such that $qm \ll 1$ ($\gamma_0 > \gamma_1 > \gamma_2  \gg 1$).
The intermediate power law asymptotic (\ref{rateintasymp3}) with exponent $1-\theta$ is again similar
to the renormalized response function due to the cascade of generations found for the self-excited
Hawkes process with just one type of events \cite{SS99,HS02,SaiSor10dissect}.
\end{enumerate}

\subsubsection{Asymptotic  and intermediate critical asymptotic inter-type activity rate $g(t):= \rho^{k,s}(t)$}

In order to obtain the time-dependence of $g(t)$, we express its Laplace
transform $\tilde{g}(u)$ given in expression  \eqref{rhomatrudiag} in a form analogous to \eqref{tilrhothruphi}:
\begin{equation}\label{tilgthruphi}
\tilde{g}(u) = \bar{R}^{k,s} \cdot {1-\varphi(u)\over 1+\gamma_0 \varphi(u)} \cdot {1\over 1+\gamma_2 \varphi(u)} ~, \qquad k\neq s~ .
\end{equation}

Three regimes can be distinguished.
\begin{enumerate}
\item {\bf Asymptotic regime of long times for $n<1$}. At long times, the
asymptotic relation \eqref{varphiasymp} holds true. Then, analogous to \eqref{tilrhouinf} and \eqref{ratettoinfty},
we obtain the following asymptotics
\begin{equation}\label{tiggtttoinf}
\begin{array}{c}\displaystyle
\tilde{g}(u) \sim \bar{R}^{k,s} \left[1- \gamma_g \beta v^\theta\right] ~, \quad \gamma_g= 1+\gamma_0+\gamma_2  \qquad u\to 0
\\[5mm] \displaystyle
\Rightarrow \quad g(t) \sim \bar{R}^{k,s} \cdot {\gamma_g \beta \theta\over \Gamma(1-\theta)}
{1 \over (\alpha t)^{1+\theta}}~ , \quad t\to \infty ~.
\end{array}
\end{equation}
This power law decay with exponent $1+\theta$, equal to the exponent of the memory kernel (\ref{fpffpowerdef}),
is characteristic of the non-critical regime  in which only a few generations of events are triggered in significant numbers.

\item   {\bf Intermediate asymptotic regime ($n \to 1$ with $q \to 0$ and $m$ large)}. Then,
$\gamma_0  \gg 1$ and $\gamma_2 \simeq 1$. This is the same second regime analyzed in subsection \ref{tjkiik}.
In this case, there is an intermediate asymptotic in the range defined by (\ref{intermediateu2}) such that
the following approximate relation holds
\begin{equation}\label{tilgtwtrhyhruphi}
\tilde{g}(u) \approx {\bar{R}^{k,s} \over \gamma_0} \cdot {1 \over  \varphi(u)} ~, \qquad k\neq s~ .
\end{equation}
The corresponding intermediate power asymptotic of $g(t):= \rho^{k,s}(t)$ is
\begin{equation}\label{rateintasymwtrh2tp3}
\rho(t) \approx {\bar{R}^{k,k}  \over \gamma_0} {\sin(\pi\theta)\over\pi}
{1 \over (\alpha t)^{1-\theta}}~ , \qquad 1 \ll \alpha t \ll \left[\gamma_0 \Gamma(1-\theta)\right]^{1/\theta} ~.
\end{equation}
This power law decay with exponent $1-\theta$ is significantly slower than the previous one with exponent $1+\theta$
and results from the proximity to the critical point $n=1$.

\item {\bf Intermediate asymptotic regime ($n \to 1$ with $q \to 0$ and $m$ small}). Then, $\gamma_2$ given in (\ref{gammasdef}) is large and there is an intermediate interval \eqref{intermediateu3} for $u$ such that, analogous to
\eqref{tilrhointermedrel}, the following approximate relation holds
\begin{equation}\label{tilgintermedrel}
\tilde{g}(u) \approx \bar{R}^{k,s} \cdot {1\over \gamma_0 \varphi(u)} \cdot {1\over \gamma_2 \varphi(u)} = \bar{R}^{k,s} \cdot {1\over \gamma_0 \gamma_2} \cdot {1\over \varphi^2(u)}~ .
\end{equation}
Using the power asymptotic \eqref{varphiasymp}, we obtain the intermediate power law
\begin{equation}
\tilde{g}(u) \approx \mathcal{G}~ v^{-2\theta} , \qquad \mathcal{G} = \bar{R}^{k,s} \cdot {1\over \gamma_0 \gamma_2 \beta} = \bar{R}^{k,s} \cdot {1\over \gamma_0 \gamma_2 \Gamma(1-\theta)}~.
\end{equation}
Accordingly, the intermediate power law asymptotic of the inter-type activity rate $g(t):= \rho^{k,s}(t)$ reads
\begin{equation}\label{crossrateintasymp}
g(t) \approx  {\bar{R}^{k,s} \over \gamma_0\gamma_2 \Gamma(2\theta) \Gamma(1-\theta)} \cdot  {1 \over (\alpha t)^{1-2\theta}}~,
~~1 \ll \alpha t \ll   \left[\gamma_2 \Gamma(1-\theta)\right]^{1/\theta}~ .
\end{equation}
This power law decay with exponent $1-2\theta$ is similar to the decay after an ``endogenous'' peak, as classified
in previous analyses of the monovariate self-excited Hawkes process \cite{HSendoexo03,Helmsorgrasso,Amazonbook1,Amazonbook2,Sorendoexo05,Cranesor08}.
Indeed, the exponent $1-2\theta$, corresponding to a very slow power law decay,
has been until now seen as the characteristic signature of
self-organized bursts of activities that are generated endogenously without the need for a
major exogenous shock. Here, we see this exponent describing the decay of the activity of
events triggered by an ``exogenous'' mother shock of a different type, in the critical regime $n \to 1$
and for weak mutual coupling $q \to 0$. It is clear that the mechanism is different from
the previously classified ``endogenous'' channel \cite{HSendoexo03,Helmsorgrasso,Amazonbook1,Amazonbook2,Sorendoexo05,Cranesor08}, involving here
an interplay between the cascade over generations and the weak mutual excitations.
\end{enumerate}

\begin{quote}
\centerline{
\includegraphics[width=11cm]{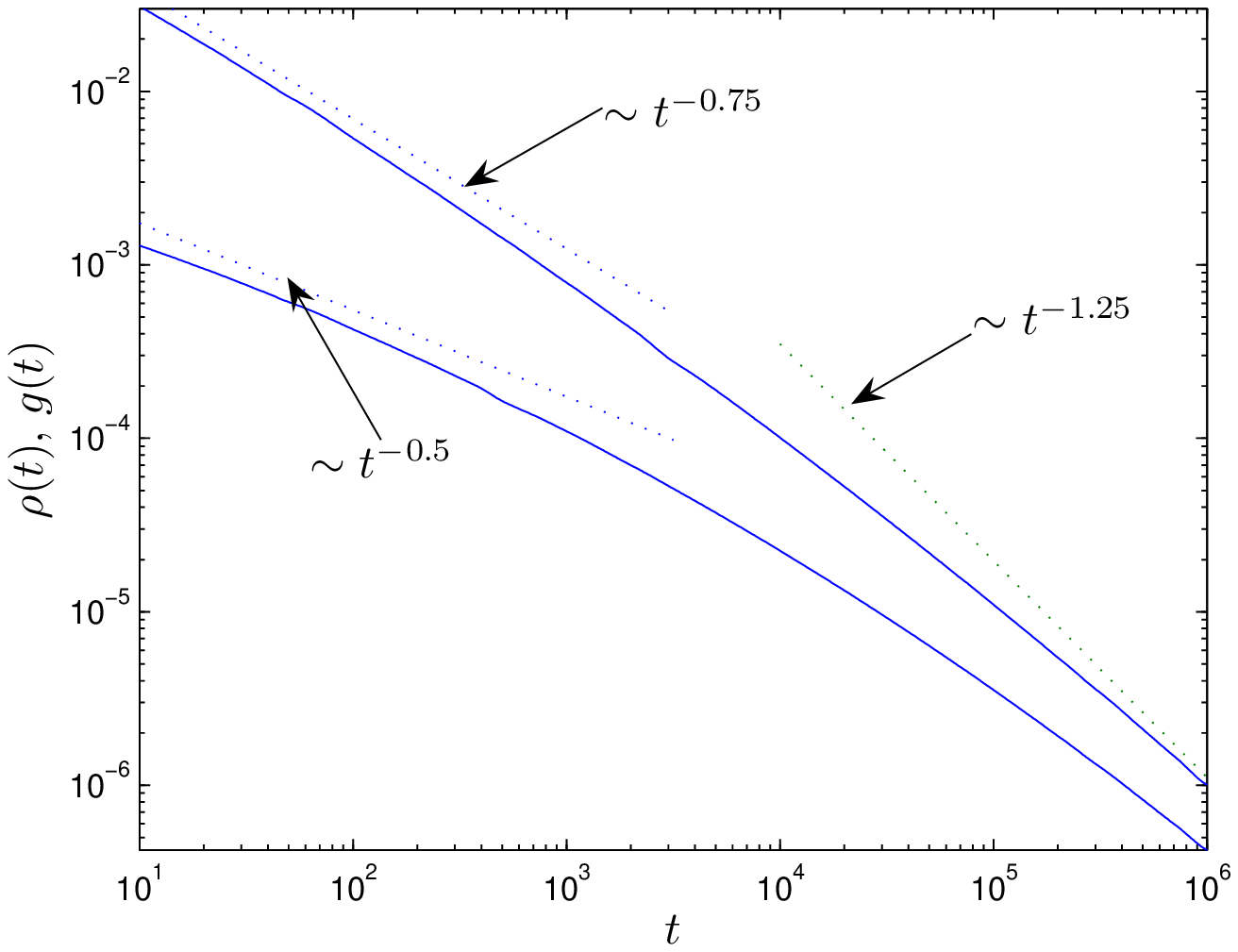}}
{\bf Fig.~3:} \small Time dependence of $\rho(t):=\rho^{k,k}(t)$ and $g(t):= \rho^{k,s}(t)$ in double
logarithmic representation
for $n=0.95$, $q=0.01$, $m=2$, and $\theta=0.25$ ($1+\theta=1.25$, $1-\theta=0.75$, $1-2\theta=0.5$).
Time is in unit of $\alpha^{-1}$. The functions $\rho(t)$ and $g(t)$ have been calculated numerically
from their complete Laplace transforms (\ref{rhomatrudiag}).
\end{quote}

Figure~3 shows the time dependence of both $\rho(t):=\rho^{k,k}(t)$ and $g(t):= \rho^{k,s}(t)$
in the regime $n \to 1$ with $q \to 0$ and $m$ small, for which there exists an intermediate
asymptotic of the third type both for $\rho(t)$ and $g(t)$. For $\rho(t)$, one can clearly observe
the intermediate power law asymptotic with exponent $1-\theta=0.75$ followed by the
final asymptotic power law with exponent $1+\theta=1.25$.  For $g(t)$, the intermediate
power law asymptotic with exponent $1-2 \theta=0.5$ is clearly observed, followed by the same
final asymptotic power law with exponent $1+\theta=1.25$.

\section{One-dimensional chain of directed triggering \label{tyjeytj}}

\subsection{Definitions}

We consider a chain of directed influences $k \to k+1$ where the events of type $k$
trigger events of both types $k$ and $k+1$ only (and not events of type $k-1$ or any
other types), and this for $k=1, 2, ...,m$.
This is captured by a form of the matrix $\hat{N}$ which has only the diagonal and
the line above the diagonal with non-zero elements.

As the simplest example, we shall study networks of mutual excitations corresponding to the following
matrix $\hat{N}$ of the mean numbers of first-generation events
\begin{equation}
\label{nmatrixablent}
\hat{N} = \left[
\begin{array}{c}
\chi ~~ \xi ~~ 0 ~~ 0 ~~ 0 ......... ~0 ...
\\
0 ~~ \chi ~~ \xi ~~ 0 ~~ 0......... ~0 ...
\\
0~~ 0 ~~ \chi ~~ \xi ~~ 0......... ~0 ...
\\
0~~ 0 ~~ 0 ~~ \chi ~~ \xi  ......... ~0 ...
\\
..................................
\end{array}
\right]
\end{equation}
where
\begin{equation}\label{ablent}
\chi={n\over 1+q} ~, \qquad \xi = {nq\over 1+q} ~.
\end{equation}

\subsection{Laplace transform of the event activities  $\rho^{k,s}(t)$ defined in \eqref{ratesdef}}

In order to derive the equations governing the rates  $\rho^{k,s}(t)$, we need to recall a result
concerning the total numbers of events $\bar{R}^{k,k}$ and $\bar{R}^{k,s}$ generated
by a given mother of type $k$. Assuming that the mother event is of type $k$,
we have \cite{SaiSormultiHanb10} $\bar{R}^{s,s} = \bar{R}^{k,s} = \bar{R}^{s,k}=0$
 for $1 \leq s <k$ and
\begin{equation}\label{rkslent}
\begin{array}{c} \displaystyle
\bar{R}^{k,k} = {\chi\over 1-\chi} = {n\over 1+q-n} ,
\\[5mm] \displaystyle
\bar{R}^{k,s} = {\xi^{s- k}\over(1-\chi)^{s- k+1}} = (1+q) {(nq)^{s- k}\over (1+q-n)^{s- k+1}} , \qquad  s > k ~ .
\end{array}
\end{equation}
Then, the Laplace transforms $\tilde{\rho}^{k,s}(u)$ \eqref{laptransfdef} of the rates $\rho^{k,s}(t)$ are
given by the right-hand sides of expressions \eqref{rkslent}, with the following substitutions
\begin{equation}
\chi \mapsto \chi\cdot \tilde{f}(u)~ , \qquad \xi \mapsto \xi \cdot \tilde{f}(u) ~.
\end{equation}
This yields
\begin{equation}\label{tilrhokslent}
\begin{array}{c} \displaystyle
\tilde{\rho}^{k,k}(u) := \tilde{\rho}(u)= {\chi\cdot \tilde{f}(u)\over 1-\chi\cdot \tilde{f}(u)}~,
\\[5mm] \displaystyle
\rho^{k,s}(u):=\tilde{g}_m(u) = {(\xi \cdot \tilde{f}(u))^{m}\over(1-\chi\cdot \tilde{f}(u))^{m+1}} , \qquad m=s-k>0 .
\end{array}
\end{equation}

\subsection{Exponential pdf $f(t)$ of triggering times of first-generation events \label{tjru}}

We use the parameterization  \eqref{ftexp} for the pdf $f(t)$, which leads after calculations to
\begin{equation}\label{rhokstlent}
\begin{array}{c} \displaystyle
\rho^{k,k}(t) := \rho(t) = \alpha \chi ~e^{-(1-\chi) \alpha t} ~,
\\[4mm] \displaystyle
\rho^{k,s}(t) := g_m(t) = {\alpha \xi \over m!} (\xi \alpha t)^{m-1} (m+\chi \alpha t) ~e^{-(1-\chi)\alpha t}~ ,~~~ m =s-k>0 ~.
\end{array}
\end{equation}
The rate $\rho^{k,k}(t) := \rho(t)$ of events of the same type as the mother decays simply as an exponential
with a characteristic decay time ${1+q \over 1+q -n} \alpha^{-1}$, which exhibits the standard
critical slowing down at the critical value of the mean branching ratio $n_c = 1+q$.
In contrast, the cross rates $\rho^{k,s}(t) := g_m(t)$ exhibit a non-monotonous behavior, which
reflects the directed nature of the mutual triggering of events of different types.
For large $m$ values, the cross-rate becomes almost symmetrical functions of time, as shown in figure~4.

\begin{quote}
\centerline{
\includegraphics[width=11cm]{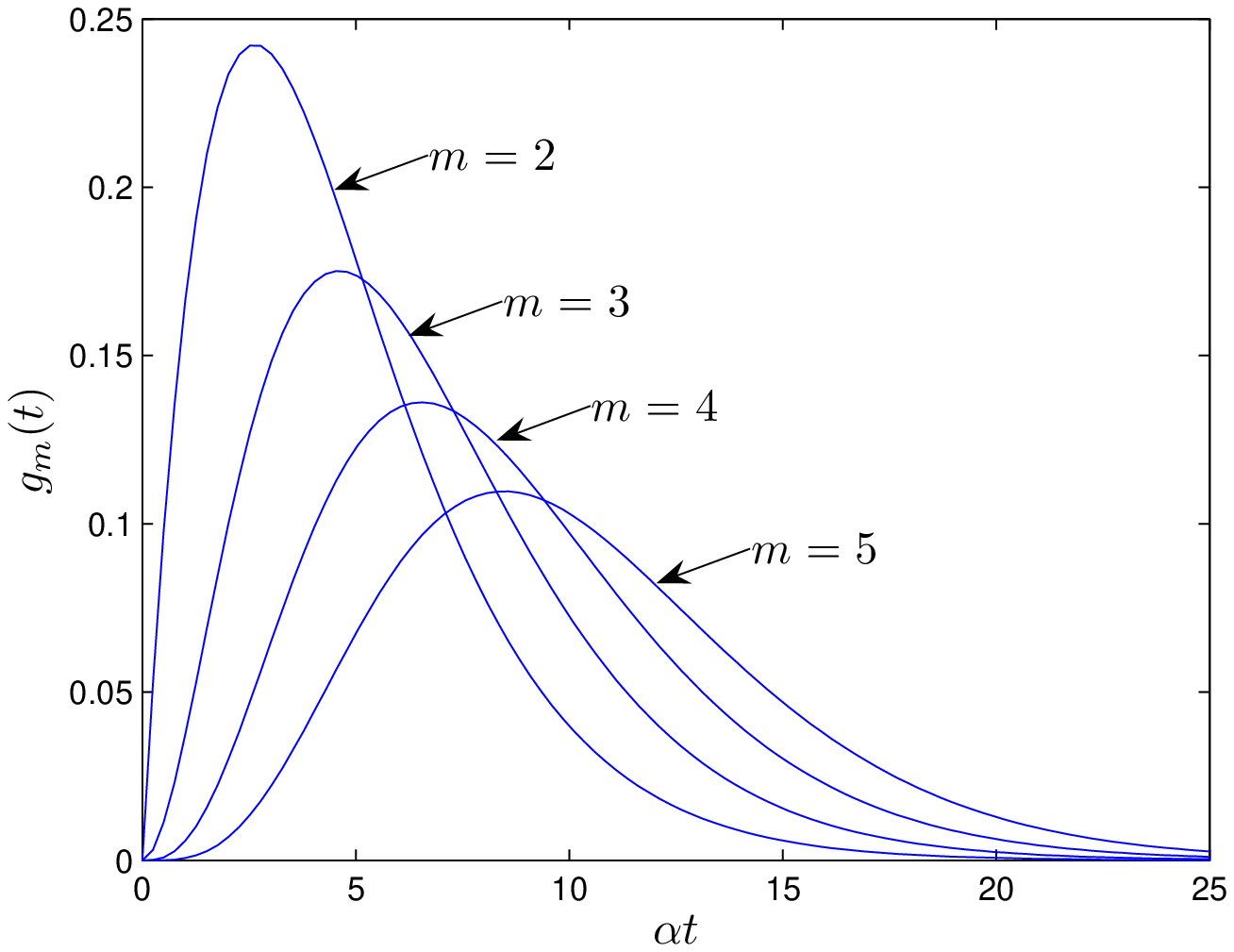}}
{\bf Fig.~4:} \small{Time dependence of the cross-rates $\rho^{k,s}(t) := g_m(t)$ given by \eqref{rhokstlent} for  $n=0.95$, $q=1$ and different $m$ values. }
\end{quote}

\subsection{Power law pdf of triggering times of first-generation events}

The time dependence of $\rho^{k,k}(t) := \rho(t)$ given by the first equation of (\ref{tilrhokslent})
when $f(t)$ is given by relation \eqref{fpffpowerdef} with Laplace transform (\ref{tildafpowlaw})
is the same as for the monovariate Hawkes process (with a single event type),
with the modification that the role of the mean branching ratio $n$
is replaced by $\chi$. The rate $\rho(t)$ exhibits an intermediate power law asymptotic with exponent $1-\theta$
up to a cross-over time $\simeq \alpha^{-1} / (1-\chi)^{1 \over \theta}$ followed by the
asymptotic power law decay with exponent $1+\theta$ corresponding to the memory kernel $f(t)$.

Interesting new regimes appear for the time dependences of the cross-rates $g_m(t)$.
We first express the Laplace transforms $\tilde{g}_m(u)$ given by \eqref{tilrhokslent} of the cross-rates
by using the auxiliary function $\varphi(u)$ defined by \eqref{varphidef}:
\begin{equation}
\label{trhkioll5i}
\tilde{g}_m(u) = \bar{R}^{k,s} \cdot {\left[1-\varphi(u)\right]^m\over \left[1+\gamma \varphi(u) \right]^{m+1}}~ ,
\end{equation}
where
\begin{equation}
\gamma = {\chi\over 1-\chi} = {n \over 1+q-n}~,
\end{equation}
and the mean number $\bar{R}^{k,s}~ (k\neq s)$ is given by expression \eqref{rkslent}.
Replacing $\varphi(u)$ by its asymptotic \eqref{varphiasymp}, we obtain the asymptotic formula
\begin{equation}\label{tilgmgamdef}
\tilde{g}_m(u) = \bar{R}^{k,s} \cdot{\left(1-\beta v^\theta\right)^m\over \left(1+\gamma \beta v^\theta \right)^{m+1}} .
\end{equation}

The long time asymptotic of  $g_m(t)$ is controlled by the behavior of $\tilde{g}_m(u)$ for $v\to 0$, whose
leading order is given by
\begin{equation}
\begin{array}{c}
\tilde{g}_m(u) \sim \bar{R}^{k,s} \left[1 - \gamma(m) v^\theta\right] ~, \qquad v\to 0 \\[3mm]
\gamma(m) = \beta (m+(m+1) \gamma)~ .
\end{array}
\end{equation}
This expression holds true for $\gamma < +\infty$, i.e., $n < 1+q$, where the upper bound $n_c=1+q$ define the
critical point.
Accordingly, the main asymptotic of the cross event rate $g_m(t)$ is
\begin{equation}
g(t) \sim \bar{R}^{k,s} \cdot {\gamma(m) \theta\over \Gamma(1-\theta)}
{1 \over (\alpha t)^{1+\theta}}~ , \quad t \to \infty~ .
\end{equation}
This recovers the usual long time power law dependence, which is determined by the memory kernel $f(t)$
of waiting times for first-generation triggering.

There is also an intermediate asymptotic regime present when $\gamma \gg 1$, i.e., $n \to 1+q$ from below.
Specifically, the intermediate asymptotic domain in the variable $v$ is defined by the interval
$(\beta \gamma)^{-1/\theta} \ll v \ll 1$,
such that $\gamma \beta v^\theta\gg 1$ while $v\ll 1$. Then,
the asymptotic relation \eqref{varphiasymp} is true, and expression \eqref{tilgmgamdef}
can be simplified into the approximate relation
\begin{equation}
\tilde{g}_m(u) \approx  \mathcal{G}_m \cdot v^{-(m+1)\theta} , \quad \mathcal{G}_m = {\bar{R}^{k,s}\over (\gamma\beta)^{m+1}}~ ,  \quad (\beta\gamma)^{-1/\theta} \ll v \ll 1~ .
\end{equation}
Accordingly, analogous to \eqref{crossrateintasymp}, we obtain
\begin{equation}\label{crossrateintmasymp}
g_m(t) \approx {\mathcal{G}_m\over \Gamma[(m+1)\theta]} \cdot
{1 \over (\alpha t)^{1-(m+1)\theta}}~ , \qquad 1 \ll \alpha t \ll (\gamma \beta)^{1/\theta} ~.
\end{equation}
This expression (\ref{crossrateintmasymp}) predicts a hierarchy of exponents $1-(m+1)\theta$ characterizing the intermediate
asymptotic power law dependence of the rates $g_m(t) := \rho^{k,s}(t)$ of events of type $s$
as a function of the distance $m=s-k$ along the space of types from the type $k$ of the initial triggering mother.
Figure~5 illustrates this prediction (\ref{crossrateintmasymp}) for $m=1;2$ with $\theta=0.25$, leading
to the two intermediate asymptotic exponents $1-2\theta=0.5$ and $1-3\theta=0.25$.

\begin{quote}
\centerline{
\includegraphics[width=11cm]{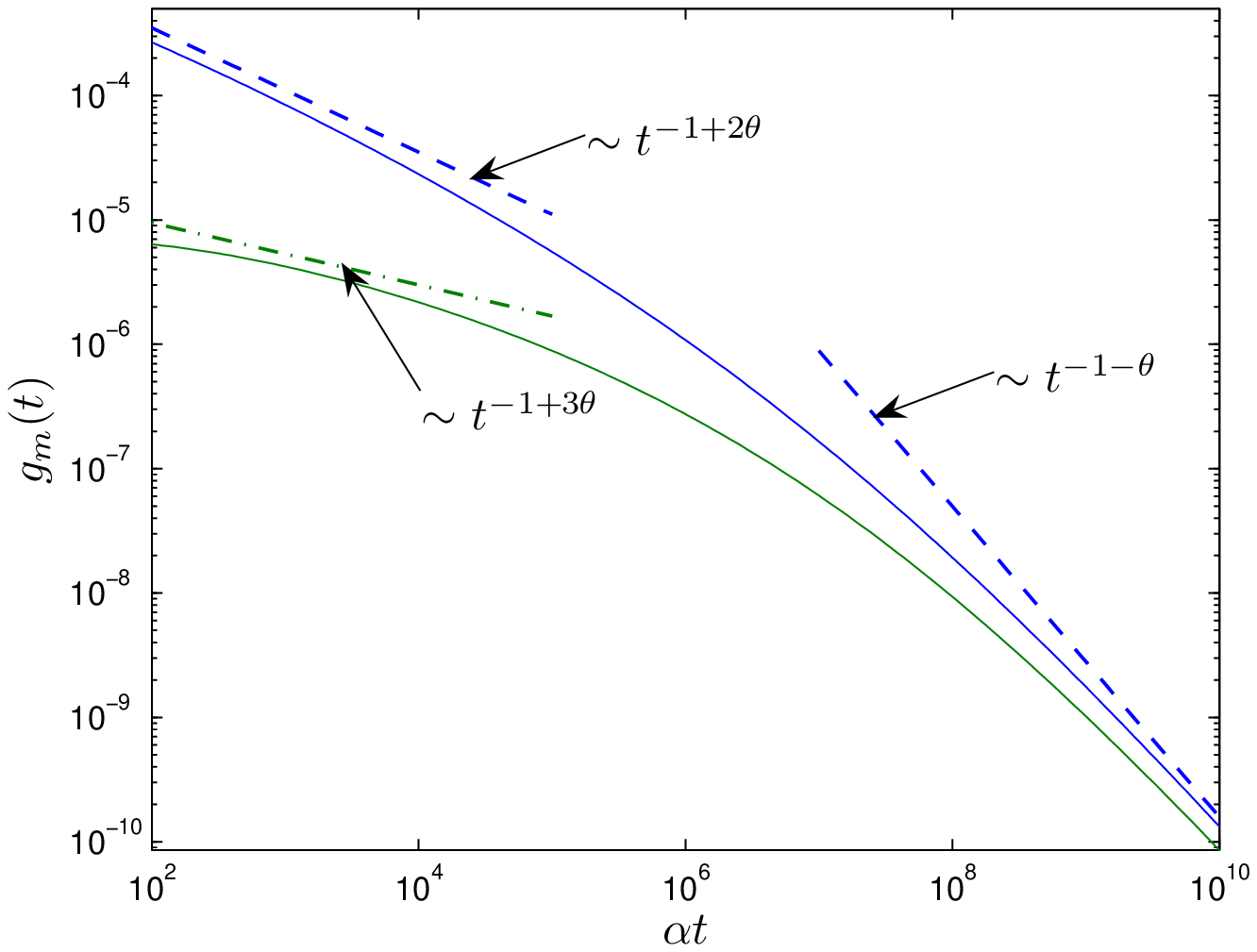}}
{\bf Fig.~5:} \small{Time dependence of the rates $g_1(t) := \rho^{k,k+1}(t)$ (upper curve)
and $g_2(t) := \rho^{k,k+2}(t)$ (lower curve) of events of type $s=k+1$ and $s=k+2$ triggered
by a mother of type $k$. The parameters are $n=0.99$, $q=0.01$ and $\theta=0.25$.
The dashed straight lines correspond to the power laws predicted in the text for the asymptotic
and corresponding intermediate asymptotic regimes. }
\end{quote}

The intermediate power law decay laws with exponents $1-(m+1)\theta$ hold only when this exponent
is positive, i.e., for $m < {1 \over \theta} -1$. To understand what happens for larger $m$'s, a more
careful analysis is required, which is presented in Appendix B, which shows that formula
(\ref{crossrateintmasymp}) still holds and predicts that $g_m(t)$ is an increasing function of time
for times $\alpha t < (\gamma \beta)^{1/\theta}$ before decreasing again with the standard
asymptotic power law $\sim 1/t^{1+\theta}$. This is summarized by figure~6, which plots
the time dependence of the rates $g_m(t) := \rho^{k,k+m}(t)$
of events of type $s=k+m$ triggered by a mother of type $k$ for $m=0$ to $5$, with $\theta=1/3$.
One can clearly observe the existence of the intermediate power asymptotics
(equation (\ref{ell}) in Appendix B and expression (\ref{crossrateintmasymp})) for different
values of $m$. When inequality (\ref{alpha}) of Appendix B holds, the intermediate
asymptotics are not decaying but growing
as a function of time, as predicted by expression (\ref{ell}) of Appendix B and (\ref{crossrateintmasymp}).

\begin{quote}
\centerline{
\includegraphics[width=13cm]{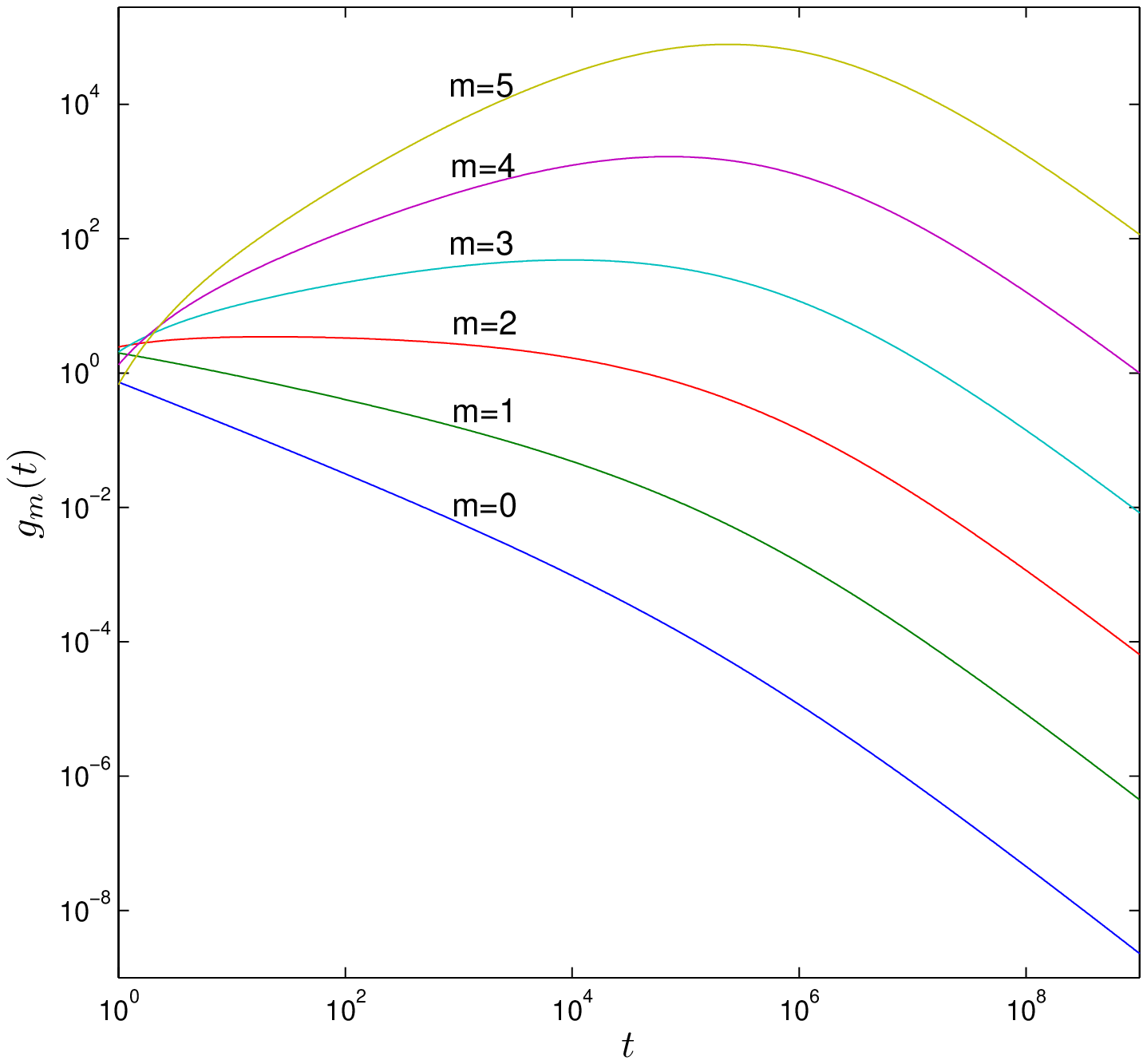}}
{\bf Fig.6:} \small{Time dependence of the rates $g_m(t) := \rho^{k,k+m}(t)$
of events of type $s=k+m$ triggered
by a mother of type $k$, for $m=0$ (same type of events as the initial triggering mother)
and $m=1, 2, 3, 4$ and $5$. Here, $\theta=1/3$ and $\gamma:=  \ln \left({1+q \over n}\right)=0.01$.
The number of summands used in the sum (\ref{c}) of Appendix B is $N=1500$.}
\end{quote}

\section{One-dimensional chain of nearest-neighbor-type triggering}

\subsection{Definitions}

A natural extension to the above one-dimensional chain of directed triggering
discussed in the previous section includes feedbacks from events
of type $k+1$ to type $k$. The example treated in the present section
corresponds to fully symmetry mutual excitations
confined to nearest neighbor in the sense of event types: $k \leftrightarrow k+1$.
Mathematically, this is described by a symmetric matrix $\hat{N}$ of the
average numbers  $n_{k,s}$ of first-generation events of different types
triggered by a mother of a fixed type.

We assume that all diagonal elements are equal to some constant $\chi$ (same self-triggering abilities)
and all off-diagnoal elements are equal to some different constant $\xi$ (same mutual triggering abilities).
The elements $n_{1,m}$ and $n_{m,1}$ are also equal to $\xi$ to close the chain of mutual excitations
between events of type $1$ and of type $m$.
Restricting to $m=6$ for illustration purpose, the corresponding matrix $\hat{N}$ reads
\begin{equation}\label{nmatrixchscircle}
\hat{N} = \left[
\begin{matrix}
\chi & \xi & 0 & 0 & 0 & \xi \\
\xi & \chi & \xi & 0 & 0 & 0 \\
0 & \xi & \chi & \xi & 0 & 0 \\
0 & 0 & \xi & \chi & \xi & 0 \\
0 & 0 & 0 & \xi & \chi & \xi
\\
\xi & 0 & 0 & 0 & \xi & \chi
\end{matrix}
\right]
\end{equation}
where
\begin{equation}\label{chccirc}
\chi={n\over 1+q}~ , \qquad \xi =  {nq\over 2 (1+q)} \quad \Rightarrow \quad \chi+ 2\xi = n~.
\end{equation}
As before, the parameter $q$ quantifies the ``strength'' of the interactions
between events of different types. Here, $n$ represents the total number of first-generation
events of all types that are generated by a given mother of fixed arbitrary type.
Figure~7 provides the geometrical sense of
matrix $\hat{N}$ \eqref{nmatrixchscircle} for $m=6$, where the circles represent the
six types of events and the arrows denote their mutual excitation influences.

\begin{quote}
\centerline{
\includegraphics[width=11cm]{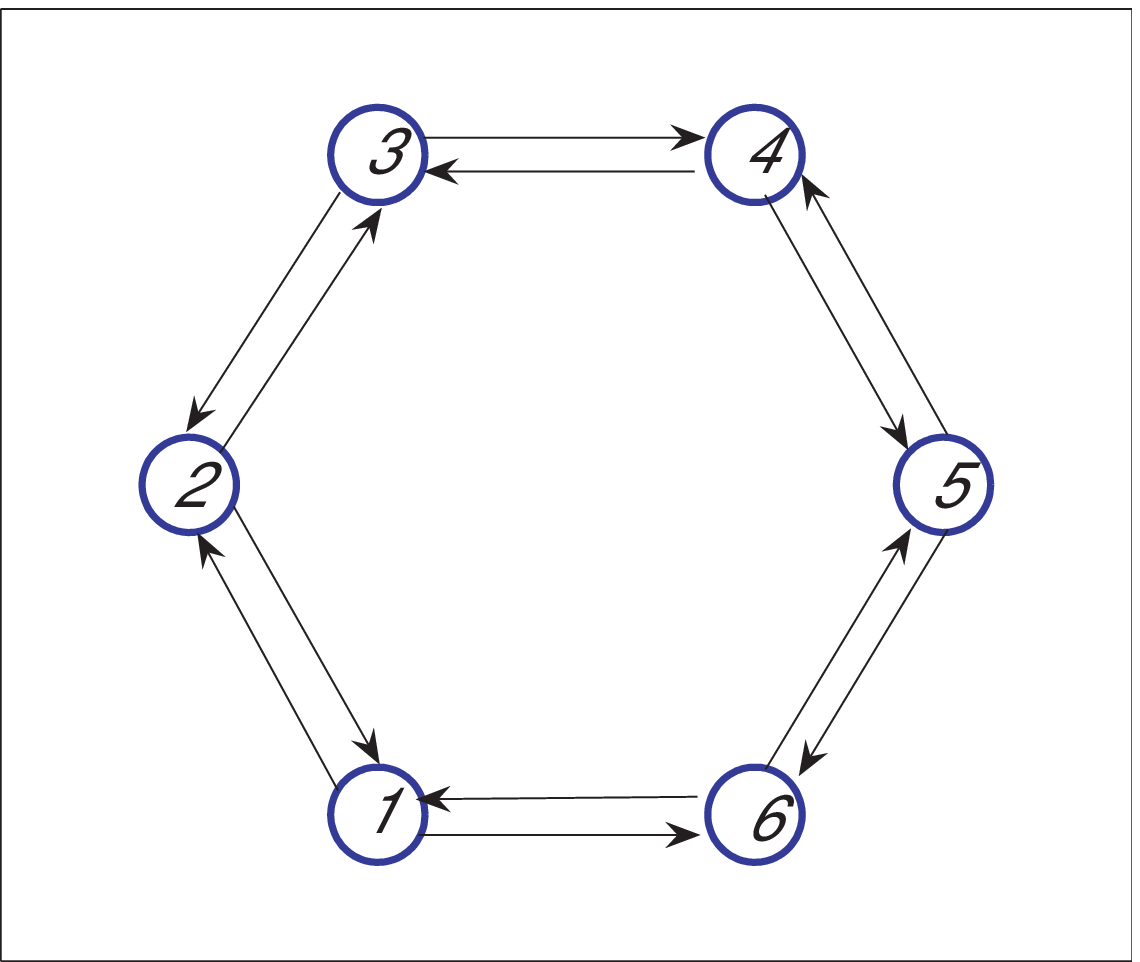}}
{\bf Fig.~7:} \small{Geometric sense of the matrix $\hat{N}$ for a one-dimensional chain of nearest-neighbor triggering in the space of event types.}
\end{quote}

\subsection{Analysis of  the event rates $\rho^{k,s}(t)$}

The Laplace transform $\tilde{\rho}^{k,k}(u)$ of the event rates  $\rho^{k,k}(t)$ \eqref{ratesdef}
of type $k$ that are triggered by a mother of the same type $k$ reads
\begin{equation}
\tilde{\rho}^{k,k}(u):=\tilde{\rho}(u)= A[n \tilde{f}(u),q]~,
\end{equation}
where \cite{SaiSormultiHanb10}
\begin{equation}
A(n,q) = \frac{4 n(1-n+q)^3-(1-n+q) (5n-2 q -2) n^2 q^2 - n^4 q^4}{ 4(1-n+q)^4- 5 (1-n+q)^2 n^2 q^2 +n^4 q^4} ~.
\end{equation}

Analogously, the Laplace transforms of the cross-rates $\tilde{\rho}^{k,s}(u)$ are
\begin{equation}
\tilde{g}_1(u) = B[n \tilde{f}(u),q] , \quad \tilde{g}_2(u) = C[n \tilde{f}(u),q] , \quad \tilde{g}_3(u) = D[n \tilde{f}(u),q] ~,
\end{equation}
where \cite{SaiSormultiHanb10}
\begin{equation}\label{abcddef}
\begin{array}{c} \displaystyle
B(n,q) = \frac{n q (1+q) (2 (1-n+q)^2  - n^2 q^2)}{ 4(1-n+q)^4- 5 (1-n+q)^2 n^2 q^2 +n^4 q^4} ~,
\\[6mm] \displaystyle
C(n,q) = \frac{n^2 q^2 (1+q)(1-n+q)}{ 4(1-n+q)^4- 5 (1-n+q)^2 n^2 q^2 +n^4 q^4} ~,
\\[6mm] \displaystyle
D(n,q) = \frac{n^3 q^3 (1+q)}{ 4(1-n+q)^4- 5 (1-n+q)^2 n^2 q^2 +n^4 q^4} ~ .
\end{array}
\end{equation}

Figure~8 shows the time dependence of $\rho(t)$, $g_1(t)$, $g_2(t)$ and $g_3(t)$ for the case
where the pdf $f(t)$ is a power law \eqref{fpffpowerdef}, for the parameters
$n=0.995, q=0.01, \theta = 0.2$. One can observe a common power law asymptotic $\sim t^{-1-\theta}=t^{-1.2}$
at large times, as well as intermediate asymptotic power laws
\begin{equation}
\begin{array}{c}
\rho(t)\sim t^{-1+\theta}=t^{-0.8} , \quad g_1(t)\sim t^{-1+2 \theta}=t^{-0.6} ~,
\\[3mm]
g_2(t)\sim t^{-1+3 \theta}=t^{-0.4} ,\quad g_3(t)\sim t^{-1+4 \theta}=t^{-0.2} ~.
\end{array}
\end{equation}

\begin{quote}
\centerline{
\includegraphics[width=13cm]{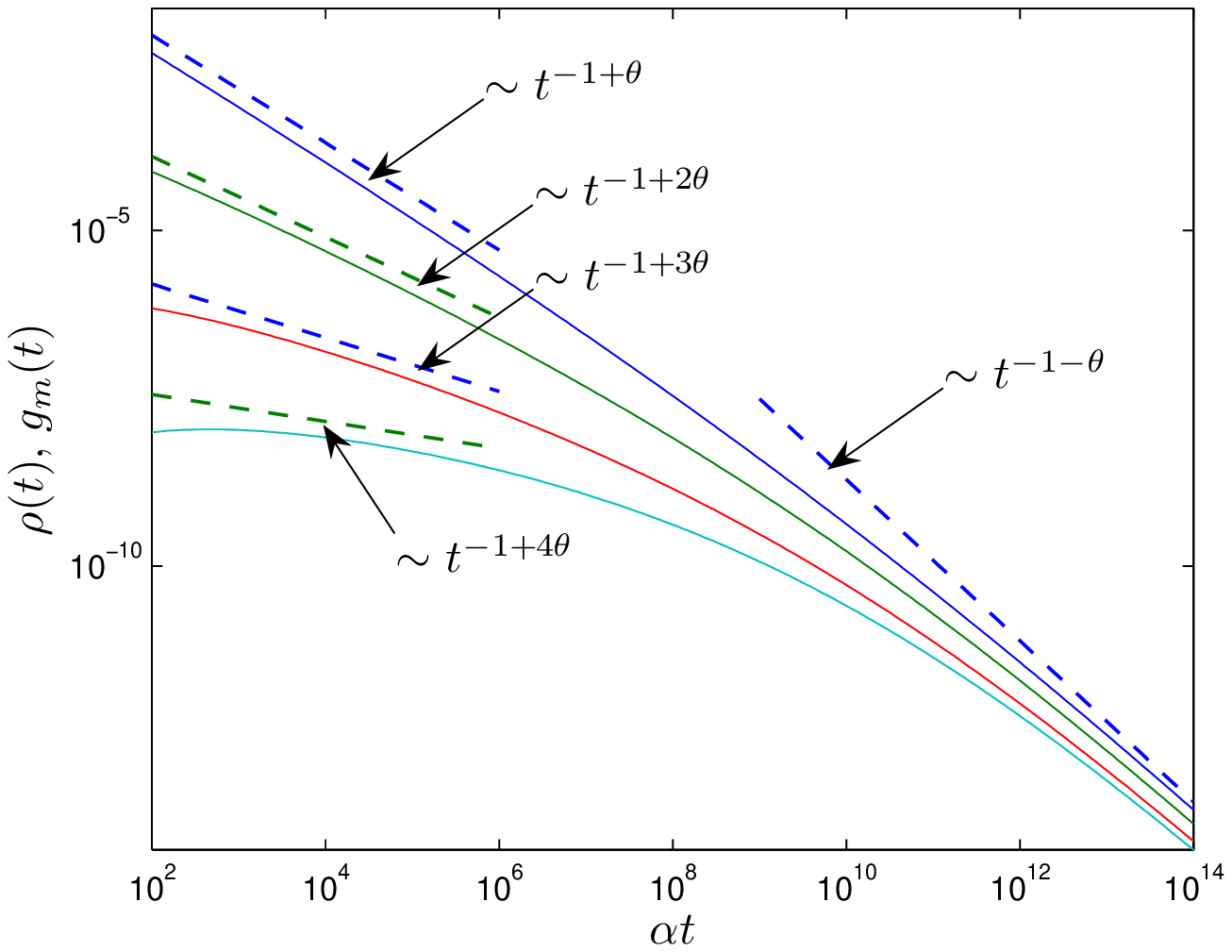}}
{\bf Fig.~8:} \small{Top to bottom: the solid lines represent the time dependence of
$\rho(t)$, $g_1(t)$, $g_2(t)$ and $g_3(t)$, in the case a one-dimensional chain of nearest-neighbor-type triggering in the space of types, with six types. The parameters are $n=0.995, q=0.01, \theta = 0.2$.
The dashed straight lines show the asymptotic and intermediate asymptotic power laws predicted in the text.}
\end{quote}

\section{Concluding remarks}

We have presented a preliminary analysis of some temporal
properties of multivariate self-excited
Hawkes conditional Poisson processes.
These processes are very interesting
candidates to model a large variety of systems with bursty events, for which past activity
triggers future activity. The term ``multivariate'' refers to the property that
events come in different types, with possibly different intra- and inter-triggering
abilities. The richness of the generated time dynamics comes from the
cascades of intermediate events of possibly different kinds, unfolding
via a kind of inter-breeding genealogy.
 We have developed the general
formalism of the multivariate generating moment function for the cumulative number
of first-generation and of all generation events triggered by a given mother event
as a function of the current time $t$. We have obtained the
general relations for the mean numbers of events triggered over all generations
by a given event as a function of time. We have applied this technical
and mathematical toolbox to several systems, characterized by different
specifications on how events of a given type may trigger events of different types.
In particular, for systems in which triggering between events
of different types proceeds through a one-dimension directed
or symmetric chain
of influence in type space, we have discovered a novel hierarchy of intermediate
asymptotic power law decays of the rate of triggered events as a function
of the distance of the events to the initial shock in the space of types.
We have been able to derive the time-dependence of the rates of events triggered from
a given shock for distributions of waiting times
of first-generation events  that have either exponential or power law tails,
for a variety of systems.
Future directions of investigations include the study of more realistic
networks in type-space and of the full distribution of even rates, beyond
the mean dynamics reported here.

\vskip 1cm
{\bf Acknowledgement}: We acknowledge financial support
from the ETH Competence Center "Coping with Crises in Complex
Socio-Economic Systems" (CCSS) through ETH Research Grant CH1-01-08-2.
This work was also partially supported by ETH Research Grant ETH-31 10-3.

\section*{Appendix A: Proof of theorem \ref{httjh6uju6}}

Substituting relation \eqref{6a} in expression (\ref{thjkuikujwthq}) leads to
\begin{equation}
\begin{array}{c}\displaystyle
A_1^k(y_1,y_2,\dots,y_m;t) = \sum_{d_1=0}^\infty \dots \sum_{d_m=0}^\infty
\\[6mm]\displaystyle
\sum_{r_1=d_1}^\infty \dots \sum_{r_m=d_m}^\infty P_k(r_1,\dots,r_m) \prod_{s=1}^m \binom{r_s}{d_s} [\mu_{k,s}(t) y_s]^{d_s} [1- \mu_{k,s}(t)]^{r_s-d_s} .
\end{array}
\label{th2gth4}
\end{equation}
Inverting the order of the summations
\begin{equation}
\sum_{d_s=0}^\infty ~\sum_{r_s=d_s}^\infty (\cdots) = \sum_{r_s=0}^\infty ~\sum_{d_s=0}^{r_s}~ ,
\end{equation}
we rewrite expression (\ref{th2gth4}) as
\begin{equation}
\begin{array}{c}\displaystyle
A_1^k(y_1,y_2,\dots,y_m;t) = \sum_{r_1=0}^\infty \dots \sum_{r_m=0}^\infty P_k(r_1,\dots,r_m)
\\[6mm]\displaystyle
\sum_{d_1=0}^{r_1} \dots \sum_{d_m=0}^{r_m}  \prod_{s=1}^m \binom{r_s}{d_s} [\mu_{k,s}(t) y_s]^{d_s} [1- \mu_{k,s}(t)]^{r_s-d_s} ~,
\end{array}
\end{equation}
or equivalently
\begin{equation}
\begin{array}{c}\displaystyle
A_1^k(y_1,y_2,\dots,y_m;t) = \sum_{r_1=0}^\infty \dots \sum_{r_m=0}^\infty P_k(r_1,\dots,r_m)
\\[6mm]\displaystyle
\prod_{s=1}^m \sum_{d_s=0}^{r_s} \binom{r_s}{d_s} [\mu_{k,s}(t) y_s]^{d_s} [1- \mu_{k,s}(t)]^{r_s-d_s}~ .
\end{array}
\end{equation}
Using the binomial formula
\begin{equation}
\sum_{d_s=0}^{r_s} \binom{r_s}{d_s} [\mu_{k,s}(t) y_s]^{d_s} [1- \mu_{k,s}(t)]^{r_s-d_s} =\left[1+\mu_{k,s} (y-1)\right]^{r_s}~ ,
\end{equation}
we obtain
\begin{equation}
A_1^k(y_1,y_2,\dots,y_m;t) =
\sum_{r_1=0}^\infty \dots \sum_{r_m=0}^\infty P_k(r_1,\dots,r_m) \prod_{s=1}^m [1+\mu_{k,s}(t)(y-1)]^{r_s}~.
\end{equation}
In view of definition \eqref{1} of the GMF $A_1^k(y_1,y_2,\dots,y_m)$, this last expression means that
\begin{equation}
A_1^k(y_1,y_2,\dots,y_m;t) = A_1^k \left[1+\mu_{k,1}(t)(y_1-1),\dots, 1+\mu_{k,m}(t)(y_m-1)\right] .
\end{equation}
Using definition \eqref{7} of the function $Q_k$, we obtain relation \eqref{6}. \hfill $\blacksquare$

\section*{Appendix B: Analysis of the behavior of $g_m(t) := \rho^{k,s}(t)$
for a one-dimensional chain of directed triggering of section \ref{tyjeytj} for $n \to 1+q$ when $1-(m+1)\theta <0$}

Let us start with expression $\tilde{g}_m(u)$ (\ref{tilrhokslent}) that we rewrite, omitting the nonessential factor $\xi^m$, as
\begin{equation}
\tilde{g}_m(u) = { \tilde{f}^{m}(u)\over[1-\chi\cdot \tilde{f}(u)]^{m+1}}~ ,~~~{\rm with}~  \chi = \frac{n}{1+q}~.
\label{68}
\end{equation}
Using the binomial formula
\begin{equation}
\frac{1}{(1-x)^{m+1}} =  \sum_{k=0}^\infty \binom{m+k}{k} x^k ~,
\end{equation}
expression (\ref{68}) becomes
\begin{equation}
\tilde{g}_m(u) = \sum_{k=0}^\infty  \binom{m+k}{k} \chi^k \tilde{f}^{m+k}(u) ~.
\label{a}
\end{equation}

As we are interesting in the case where the pdf $f(t)$ has the power asymptotic
$f(t) \sim 1/t^{1+\theta}$, with $0<\theta< 1$, it is convenient to use for $f(t)$ one special representative
of the functions presenting this asymptotic power law behavior, namely the one-sided
L\'evy stable distribution of order $\theta$ \cite{Zolotarev}, that we refer to as $f_\theta(t)$. Its Laplace transform is
\begin{equation}
\tilde{f}_\theta(u) = e^{-u^\theta} , \qquad 0<\theta< 1~ .
\end{equation}
Accordingly, relation (\ref{a}) takes the form
\begin{equation}
\tilde{g}_m(u) = \sum_{k=0}^\infty  \binom{m+k}{k} \chi^k \cdot e^{-(m+k) u^\theta} ~.
\label{ytjukii}
\end{equation}
Taking the inverse Laplace transform of (\ref{ytjukii}) provides us with the exact expression
\begin{equation}
g_m(t) = \sum_{k=0}^\infty  \binom{m+k}{k}  \cdot \frac{\chi^k}{(m+k)^{1/\theta}} \cdot
f_\theta\left(\frac{t}{(m+k)^{1/\theta}}\right)~ .
\label{b}
\end{equation}

In order to analyze (\ref{b}), it is convenient to rewrite it as
\begin{equation}
g_m(t) = \frac{\theta}{t^{1+\theta}} ~ \sum_{k=0}^\infty S_m(k+m) ~ Q_\theta\left( \frac{m+k}{t^\theta}\right)  e^{-\gamma k} ~,
\label{c}
\end{equation}
where
\begin{equation}
\gamma = \ln\left(\frac{1}{\chi}\right) = \ln \left({1+q \over n}\right)  > 0 ~~ {\rm for} ~n< 1+q~.
\end{equation}
We have defined the functions
\begin{equation}
S_m(x) := x \binom{x}{m} , \qquad x\geqslant m~ ,
\end{equation}
and
\begin{equation}
Q_\theta(x) := \frac{1}{\theta x^{1+1/\theta}} f_\theta\left(\frac{1}{x^{1/\theta}}\right) , \qquad x\geqslant 0 ~,
\end{equation}
such that
\begin{equation}
\frac{1}{x^{1/\theta}} f_\theta\left(\frac{t}{x^{1/\theta}}\right) = \frac{x\,\theta}{t^{\theta+1}} Q_\theta\left(\frac{x}{t^\theta}\right) ~.
\end{equation}

In order to extract the relevant information from expression (\ref{c}) for $g_m(t)$, we need
to discuss some properties of the two functions $S_m(x)$ and $Q_\theta(x)$.
\begin{itemize}
\item[] {\bf Properties of the function $S_m(x)$}.
The function $S_m(x)$ is a finite sum of power functions of the argument $x$
\begin{equation}
S_m(x) = \sum_{r=1}^{m} a_{r,m} ~ x^{r+1}~ .
\label{d}
\end{equation}
In particular,
\begin{equation}
\begin{array}{c}\displaystyle
S_0(x) = x , \qquad S_1(x) = x^2 , \qquad S_2(x) = \frac{1}{2}~ x^3 - \frac{1}{2}~ x^2~ ,
\\[4mm] \displaystyle
S_3(x) = \frac{1}{6}~ x^4 - \frac{1}{2}~ x^3 + \frac{1}{3}~ x^2~ ,
\\[4mm] \displaystyle
S_4(x) = \frac{1}{24}~ x^5 - \frac{1}{4}~ x^4 + \frac{11}{24}~ x^3 - \frac{1}{4}~ x^2~ .
\end{array}
\label{e}
\end{equation}

\item[] {\bf Properties of the function $Q_\theta(x)$}.
For $\theta\in(0,1/2)$, this function is at least exponentially decaying with increasing $x$.
Accordingly, its moments of any order $r>0$ are finite and given by
\begin{equation}
M(r):= \int_0^\infty x^r ~ Q_\theta(x) dx = \frac{\Gamma(r+1)}{\Gamma(r\theta+1)} ~.
\label{f}
\end{equation}
Moreover, the value of $Q_\theta(x)$ at $x=0$ is equal to
\begin{equation}
Q_\theta(x=0) = \frac{1}{\Gamma(1-\theta)} ~.
\label{g}
\end{equation}
For the particular cases $\theta=1/2$ and $\theta=1/3$, the function $Q_\theta(x)$ can be expressed
in explicit form:
\begin{equation}
Q_{1/2}(x) = \frac{1}{\sqrt{\pi}} \exp\left(-\frac{x^2}{4}\right) , \qquad Q_{1/3}(x) = \sqrt[3]{9} \cdot \text{Ai}\left(\frac{x}{\sqrt[3]{3}}\right) ~.
\label{h}
\end{equation}
\end{itemize}

We study the behavior of $g_m(t)$ given by expression (\ref{c}) for $\gamma m \ll 1,  t^\theta \gg 1$ and $m~t^{-\theta}\ll 1$.
To leading order and without essential error,  we may replace the discrete sum (\ref{c}) by the continuous integral
\begin{equation}
g_m(t) \simeq \frac{\theta}{t^{1+\theta}} \int_m^\infty S_m(x) ~Q\left(\frac{x}{t^{\theta}}\right) e^{-\gamma x} dx~ .
\label{eytjuht}
\end{equation}
Using the following change of variable of integration
\begin{equation}
x \quad \mapsto \quad y = \frac{x}{t^\theta}  \quad \leftrightarrow \quad x = t^\theta ~ y~,
\end{equation}
we obtain
\begin{equation}
g_m(t) \simeq \frac{\theta}{t} \int_0^\infty S_m(t^\theta y) Q_\theta(y) e^{-\gamma t^\theta y} dy~ .
\end{equation}
Using relation (\ref{d}), this yields
\begin{equation}
g_m(t) \simeq \frac{\theta}{t} \sum_{r=1}^m a_{r,m} t^{(r+1)\theta} G_{r,m}(t)~ ,
\label{i}
\end{equation}
where
\begin{equation}
G_{r,m}(t) = \int_0^\infty y^{r+1} Q_\theta(y) e^{-\gamma t^\theta y} dy ~.
\label{k}
\end{equation}

The intermediate asymptotic regime corresponds to the time domain $\gamma t^\theta\ll 1$ with $t^\theta\gg 1$.
In this case, the exponential in the integral (\ref{k}) can be replaced by unity. Using relations (\ref{f}), we obtain that
\begin{equation}
G_{r,m} = \int_0^\infty y^{r+1} Q_\theta(y) dy =\frac{\Gamma(r+2)}{\Gamma[(r+1)\theta+1]}~ , \qquad \gamma t^\theta \ll 1 ~,
\end{equation}
is time independent. Accordingly, the mean rate $g_m(t)$ (\ref{i}) is found as the sum of power law functions
\begin{equation}
g_m(t)\simeq \theta \cdot t^{(m+1)\theta-1} \cdot \sum_{r=1}^m a_{r,m}  G_{r,m} \cdot t^{(r-m)\theta}~.
\end{equation}
Taking into account that we consider the case $t^\theta\gg 1$ (that allowed us to use the integral approximation
(\ref{eytjuht})), we obtain the sought power law intermediate asymptotic
\begin{equation}
g_m(t) \simeq \frac{\theta (m+1)}{\Gamma[(m+1)\theta+1]} \cdot t^{(m+1)\theta-1} \sim t^{(m+1)\theta-1}, \qquad 1 \ll t \ll \gamma^{-1/\theta} ~.
\label{ell}
\end{equation}
This recovers the result (\ref{crossrateintmasymp}) presented in the main text. In addition, it
makes more precise what happens for
\begin{equation}
(m+1)\theta>1 ~.
\label{alpha}
\end{equation}
In this case, $g_m(t)$ starts as a growing function of $t$ up to $t \simeq \gamma^{-1/\theta}$.
This retrieves the same qualitative behavior found when $f(t)$ is an exponential, which has been
analyzed in subsection \ref{tjru} and represented in figure~4.

Of course, at times $t \gg \gamma^{-1/\theta}$, this growth is replaced by the standard power law
decay $\sim 1/t^{1+\theta}$. Indeed, for $\gamma t^\theta\gg 1$, the integral (\ref{k}) is approximately equal to
\begin{equation}
G_{r,m}(t) \simeq  Q_\theta(0) \int_0^\infty y^{r+1} e^{-\gamma t\theta y} dy = \frac{(r+1)!}{\Gamma[1-\theta]} \gamma^{-r-2} t^{-(r+2)\theta}~ .
\end{equation}
Substituting this relation into (\ref{i}) yields
\begin{equation}
g_m(t) \simeq \frac{\theta}{t^{1+\theta}} \sum_{r=1}^m a_{r,m} \frac{(r+1)!}{\Gamma[1-\theta]} \gamma^{-r-1}
\sim {1 \over t^{1+\theta}} ~,  \qquad t\gg \gamma^{-1/\theta}~ .
 \label{n}
\end{equation}

Figure~6 in the text sums up these results by plotting the time dependence of the rates $g_m(t) := \rho^{k,k+m}(t)$
of events of type $s=k+m$ triggered by a mother of type $k$ for $m=0$ to $5$.
One can clearly observe the existence of the intermediate power asymptotics (\ref{ell}) for different
values of $m$. When inequality (\ref{alpha}) holds, the intermediate asymptotics are not decaying but growing
as a function of time, as predicted by expression (\ref{ell}).

\end{document}